\documentclass[10pt]{article}
\usepackage{amssymb,amsmath}
\usepackage{amsfonts}
\usepackage{eufrak}
\usepackage{graphicx}
\begin{document}

\twocolumn[\begin{center}
{\large {\bf  Numerical Solutions of ideal two-fluid equations very closed to the event horizon of Schwarzschild black hole}}\\
\vspace{2cm}
M. Atiqur Rahman\\
{\it Department of Applied Mathematics,\\ Rajshahi University,
Rajshahi - 6205, Bangladesh}

\vspace{1cm}
\centerline{\bf Abstract}
\baselineskip=18pt
\bigskip
\parbox{15.5cm}{The $3+1$ formalism of Thorne, Price and Macdonald has been used to derive the linear two-fluid equations describing transverse and longitudinal waves propagating in the two-fluid ideal collisionless plasmas surrounding a Schwarzschild black hole. The plasma is assumed to be falling in radial direction toward the event horizon. The relativistic two-fluid equations have been reformulate, in analogy with the special relativistic formulation as  explained in an earlier paper, to take account of relativistic effects due to the event horizon. Here a WKB approximation is used to derive the local dispersion relation for these waves and solved numerically for the wave number $k$.\\
{E-mail: $atirubd@yahoo.com$}\\
{\it PACS}: 95.30.Qd, 95.30.Sf, 97.60.Lf\\
{\it Keywords}: Transverse and Longitudinal Wave Modes, Schwarzschild Black Hole}\\
\end{center}
]\section{Introduction}\label{sec1}
Black holes belong to the most fascinating objects predicted by Einstein\rq s theory of gravitation and are still mysterious \cite{one}. Physicists are grappling the theory of black holes, while astronomers are searching for real-life examples of black holes in the universe \cite{two}. Recent observations of nuclei of galaxies have revealed large mass concentrations over relatively small volumes \cite{three,four,five,six,seven,eight,nine,ten}. The only plausible explanation for high compactness of the mass distribution is that galactic nuclei, including the Milky Way, contain a massive or a suppermassive $(10^6-10^9 M_\odot)$ black hole. Classical black holes do not emit electromagnetic waves and so it is not possible to observe them directly. Their presence is inferred by observing physical processes occurring in the plasma situated around their event horizon. Perturbed black holes do emit  gravitational waves thus making it possible to observe them directly.

Within $3R_s$ (3 Schwarzschild radii) it is possible to have plasma \cite{eleven,twelve,thirteen,fourteen}. The theory of general relativity and its application to the plasma close to the black hole horizon have remained esoteric, and little concrete astrophysical impact has been felt. In recent year plasma equations and general relativity are usually considered together. The Coulomb potential of charge particles due to coupling is much stronger than the gravitational potential and are often neglected in the Newtonian approximation. But the mean gravitational field for certain astronomical objects like galactic nuclei or black holes may be strong and the observation of magnetic fields indicates that a combination of general relativity and plasma physics at least on the level of a fluid description is appropriate. The plasma in the black hole environment may act as a fluid and black holes greatly affect the surrounding plasma medium (which is highly magnetized) with their enormous gravitational fields. Hence, it is therefore of interest to formulate  the plasma physics problem in context of general relativity.

A covariant formulation of the theory based on the fluid equations of general relativity and Maxwell's equations in curved spacetime has so far proved unproductive because of the curvature of four-dimensional spacetime in the region surrounding a black hole. MacDonald and Thorne \cite{fifteen,sixteen} have introduced Maxwell\rq s equations in 3+1 coordinates, which provides a foundation for formulation of a general relativistic (GR) set of plasma physics equations in the strong gravitational field of black hole and the \lq\lq membrane paradigm\rq\rq \cite{seventeen} is a good example of a formalism which has been developed for this purpose.

The membrane paradigm is mathematically equivalent to the standard, full general relativistic theory of black holes, so far as all physics outside the horizon is taken into account in which the black holes event horizon are replaced with a surrogate \lq\lq stretched horizon\rq\rq viewed as a two-dimensional membrane that resides in three-dimensional space and evolves in response to driving forces from the external universe. But the formulation of all physics in this region turns out to be very much simpler than it would be using the standard covariant approach of general relativity. In the 3+1 formulation, work connected with black holes has been facilitated by replacing the hole's event horizon with a membrane endowed with electric charge, electrical conductivity, and finite temperature and entropy so that the physics outside the event horizon turns out to be very much simpler than it would be using the standard covariant approach of general relativity.

Exploiting the 3+1 formalism, a lot of works have been carried out. The study of plasma wave in the presence of strong gravitational fields using the $3+1$ approach is still in its early stages. Zhang \cite{eighteen,nineteen} has considered the care of ideal magneto hydrodynamics waves near a Kerr black hole, accreting for the effects of the holes angular momentum but ignoring the effects due to the black hole horizon. Holcomb and Tajima \cite{twenty}, Holcomb \cite{twenty one}, and Dettmann et. al. \cite{twenty two} have considered some properties of wave propagation in a Friedmann universe. Daniel and Tajima \cite{twenty three} studied the physics of high frequency electromagnetic waves in a strong Schwarzschild plasma.

The most interesting plasma phenomena due to involvement of collective effects are strongly nonlinear even in the Newtonian or special relativistic case. The addition of gravity can be expected to cause even more highly nonlinear and violent phenomena. Existing plasma theories and computational codes are based on Newtonian physics, perhaps accounting for special relativity. Sakai and Kawata (SK) \cite{twenty four} have developed the linearized treatment of plasma waves using special relativistic formulation. Connecting this work Buzzi, Hines, and Treumann (BHT) \cite{twenty five, twenty six} have developed  general relativistic two-fluid plasma waves around Schwarzschild black hole using local approximation. Using local approximation near the event horizon of different nonrotating black holes some research have been made \cite{twenty seven, twenty eight, twenty nine, thirty}. Recently, in Ref. \cite{thirty one} the two-fluid equations for transverse and longitudinal waves are simplified using action principle developed by Heintzmann and Novello \cite{thirty two} and solved using the analytical method developed by Mikhailovskii \cite{thirty three}. In this paper, the dispersion relations for transverse and longitudinal waves are solved numerically using WKB approximation.

The principal objective of the work presented in this paper is to make an initial attempt to be the discovery of an instability caused by the general relativistic term in the local dispersion relations for transverse (electromagnetic) and longitudinal (electrostatic) waves using WKB approximation. Such an investigation of wave propagation in a general relativistic two-fluid plasmas near a black hole is important for an understanding of plasma processes. That is, what happens when the plasma are assumed to be infalling onto the black hole.

In the present paper Sec. \ref{sec2} summarizes the $3+1$ formulation of Schwarzschild black hole Spacetime. The 3+1 formulism of the two-fluid plasmas expressing continuity, the conservation of energy and momentum, and Maxwell\rq s equations in Schwarzschild coordinates are presented in Sec. \ref{sec3}. The transformation of Schwarzschild metric in Rindler coordinates is also described here. By assuming the two-fluid plasma falling toward the radial direction of the event horizon, the transverse and longitudinal components of the two-fluid equations are separated by introducing a new complex transverse fields and velocities using Cartesian coordinates in Sec. \ref{sec4}. The linearized two-fluid equations for the transverse and longitudinal waves are derived by considering a small perturbation to the fields and fluid parameters in Sec. \ref{sec5}. In Sec. \ref{sec6} we discuss the way in which the unperturbed fields and fluid parameters and their derivatives with respect to $z$ depend on the surface gravity and the infall velocity from the black hole horizon. Thin layers are considered to discover the insstability of two fluid plasma wave near the horizon, each with its own appropriate mean value of the lapse function and the two-fluid equations are simplifying using WKB approximation in Sec. \ref{sec7}. In Sec. \ref{sec8} we explain the dispersion relations for the transverse and longitudinal waves. The numerical method for solving the two-fluid equations is discussed in Sec. \ref{sec9}. The numerical results for the Alfv\'en, high frequency electromagnetic, and longitudinal waves are also discussed in this section.  Finally, in Sec. \ref{sec10} we present our remarks. Here, we use units in which $G=c=k_B=1$.

\section{3+1 Formalism of\\ Schwarzschild Spacetime}\label{sec2}
As mentioned in the previous section, our work presented in this paper is based on the 3+1 formulation of general relativity developed by Thorne, Price, and Macdonald (TPM) \cite{fifteen,sixteen,seventeen}. Actually the $3+1$ approach was originally developed in 1962 by Arnowitt, Deser, and Misner \cite{thirty four} to study the quantization of the gravitational field. Since then, their formulation has most been applied in studying numerical relativity \cite{thirty five}. TPM extended the $3+1$ formalism to include electromagnetism and applied it to study electromagnetic effects near the Kerr black hole. So their work has opened up many possibilities for studying relativistic effects on plasma around black hole horizon in the electromagnetic window. The basic concept behind the $3+1$ formulation of general relativity is to select a preferred set of spacelike hypersurfaces which form the level surfaces of a congruence of timelike curves. A particular set of these hypersurfaces constitutes a time slicing of spacetime. The hypersurfaces considered here are of constant universal time $t$. In the $3+1$ formulation, the Schwarzschild metric is given by
\begin{align}
&ds^2={\rm g}_{\mu \nu }dx^\mu dx^\nu\nonumber\\
&=-\left(1-\frac{2M}{r}\right)dt^2+\frac{1}{(1-2M/r)}dr^2+r^2d\Omega^2,\label{eq1}
\end{align}
where $d\Omega^2=d\theta ^2+{\rm sin}^2\theta d\varphi ^2$. The components $x^\mu $ denote spacetime coordinates and $\mu , \nu =0, 1, 2, 3$. The hypersurfaces of constant universal time $ t$ define an absolute three-dimensional space described by the metric
\begin{equation}
ds^2={\rm g}_{ij}dx^idx^j=\frac{1}{(1-2M/r)}dr^2+r^2d\Omega^2\label{eq2}.
\end{equation}
The indices $i, j$ range over $1,2,3$ and refer to coordinates in absolute space. The curl and divergence operators in the 3+1 set of equations are covariant and can be derived from the above metric. The purpose of choosing such an absolute three-dimensional space is to isolate the effect of the gravitational field from the choice of local Cartesian coordinates.

We consider a set of fiducial observers (FIDOs), i.e. the observers remaining at rest with respect to this absolute space, measure their proper time $\tau $ using clocks that they carry with them and make local measurements of all physical quantities. Then their all measured quantities are defined as FIDO locally measured quantities and all rates measured by them are measured using FIDO proper time. For the absolute matric space given in Eq. (\ref{eq2}), FIDOs use a local Cartesian coordinate system with unit basis vectors tangent to the coordinate lines as
\begin{equation}
{\bf e}_{\hat r}=(1-2M/r)^{1/2}\frac{\partial }{\partial r},{\bf e}_{\hat \theta }=\frac{1}{r}\frac{\partial }{\partial \theta },{\bf e}_{\hat \varphi }=\frac{1}{r{\rm sin}\theta }\frac{\partial }{\partial \varphi}\label{eq3}.
\end{equation}
For a spacetime viewpoint rather than a $3+1$ split of spacetime, the set of orthonormal vectors also includes the basis vector for the time coordinate given by
\begin{equation}
{\bf e}_{\hat 0}=\frac{d}{d\tau }=\frac{1}{\alpha }\frac{\partial }{\partial t}\label{eq4},
\end{equation}
where $\alpha $ is the lapse function (or redshift factor) defined by
\begin{equation}
\alpha (r)\equiv \frac{d\tau }{dt}=\left(1-\frac{2M}{r}\right)^{1/2}\label{eq5}.
\end{equation}
The lapse function $\alpha$  has an important plays the role of a gravitational potential and thereby governs the ticking rates of clocks and redshifts. We can calculate the gravitational acceleration felt by a FIDO from the lapse function as follows \cite{twenty five, twenty six}:
\begin{equation}
{\bf a}=-\nabla {\rm ln}\alpha =-\frac{1}{\alpha }{\frac{M}{r^2}}{\bf e}_{\hat r}\label{eq6},
\end{equation}
Equation (\ref{eq6}) shows that far from the black hole event horizon i.e., when $\alpha \rightarrow 1$, the gravitational acceleration becomes weak and approaches the Newtonian value for flat spacetime. However, near the horizon, the gravitational acceleration approaches infinity as $\alpha \rightarrow 0$. The rate of change of any scalar physical quantity or any three-dimensional vector or tensor, as measured by a FIDO, is defined by the derivative
\begin{equation}
\frac{D}{D\tau }\equiv \left(\frac{1}{\alpha }\frac{\partial }{\partial t}+{\bf v}\cdot \nabla \right)\label{eq7},
\end{equation}
$\bf v$ being the velocity of a fluid as measured locally by a FIDO. Since all the quantities are measured locally by the FIDO, all the vector quantities are neither covariant nor contravariant. The derivation of the gravitational accelerations for different spacetime matrices are detailed in TPM \cite{fifteen,sixteen,seventeen} and BHT et al \cite{twenty five, twenty six}.

\section{Two-fluid Equations\\ in 3+1 Formalism}\label{sec3}
The two-fluid plasmas situated around the black hole horizon are considered as ideal fluids having two components either electron-positron or electron-ion to describe the two-fluid equations for the continuity, conservation of energy and momentum, and Maxwell's equations in $3+1$ formalism. These type of fluid components are chosen because the dispersion relations that result from the following investigation are valid for either one of the two fluids since no assumption can be made upon the mass, number density, pressure or temperature of the fluids. BHT \cite{twenty five, twenty six} have derived the equation of continuity and the conservation of energy and momentum and Maxwell\rq s equations in 3+1 split of spacetime. The derivation of the equations required for all wave disturbances in the plasma including the general relativistic effects are detailed here. The reader is referred to TPM  \cite{fifteen,sixteen,seventeen} for this material. In the BHT works \cite{twenty five, twenty six} , the equation of continuity and Maxwell's equations coupling with the two-fluid having velocity ${\bf v}_s$, mass $m_s$, number density $n_s$, and charge $q_s$ for electromagnetic fields ($\bf E$ and $\bf B$) are given by
\begin{equation}
\frac{\partial }{\partial t}(\gamma _sn_s)+\nabla \cdot (\alpha \gamma _sn_s{\bf v}_s)=0\label{eq8},
\end{equation}
and
\begin{eqnarray}
\nabla \cdot {\bf B}&=&0,\label{eq9}\\
\nabla \cdot {\bf E}&=&4\pi \sigma ,\label{eq10}\\
\frac{\partial {\bf B}}{\partial t}&=&-\nabla \times (\alpha {\bf E}),\label{eq11}\\
\frac{\partial {\bf E}}{\partial t}&=&\nabla \times (\alpha {\bf B})-4\pi \alpha {\bf J},\label{eq12}
\end{eqnarray}
with the charge and current densities defined as
\begin{equation}
\sigma =\sum_s\gamma _sq_sn_s,\hspace{1.2cm}{\bf J}=\sum_s\gamma _sq_sn_s{\bf v}_s\label{eq13},
\end{equation}
where $\gamma_s$ is the relativistic Lorentz factor and $s$ is $1$ for electrons and $2$ for positrons (or ions). The lapse function $\alpha$ signifies the general relativistic effect around a Schwarzschild black hole. The equations for the conservation of energy and momentum, as derived by BHT \cite{twenty five, twenty six}, using above Maxwell's equations for electromagnetic fields are given, respectively, for  each species $s$ by
\begin{equation}
\frac{1}{\alpha }\frac{\partial }{\partial t}\epsilon _s=-\nabla \cdot {\bf S}_s+2{\bf a}\cdot {\bf S}_s,\label{eq14}
\end{equation}
and
\begin{equation}
\frac{1}{\alpha }\frac{\partial }{\partial t}{\bf S}_s=\epsilon _s{\bf a}-\frac{1}{\alpha }\nabla \cdot (\alpha {\stackrel{\longleftrightarrow }{\bf W}}_s)\label{eq15}.
\end{equation}
Where the energy density $\epsilon _s$, momentum density ${\bf S}_s$, and stress-energy tensor $W_s^{jk}$ for the electromagnetic field have been taken as
\begin{align}
&\epsilon _s=\frac{1}{8\pi }({\bf E}^2+{\bf B}^2),\quad{\bf S}_s=\frac{1}{4\pi }{\bf E}\times {\bf B},\nonumber\\
&W_s^{jk}=\frac{1}{8\pi }({\bf E}^2+{\bf B}^2)g^{jk}-\frac{1}{4\pi }(E^jE^k+B^jB^k).\label{eq16}
\end{align}
For a ideal relativistic fluid of species $s$ in three-dimensions, the energy density $\epsilon _s$, momentum density ${\bf S}_s$, and stress-energy tensor $W_s^{jk}$ corresponding to the electromagnetic field quantities given in Eq. (\ref{eq16}) are
\begin{eqnarray}
&&\epsilon _s=\gamma _s^2(\varepsilon _s+P_s{\bf v}_s^2),\quad{\bf S}_s=\gamma _s^2(\varepsilon _s+P_s){\bf v}_s, \nonumber\\
&&W_s^{ij}=\gamma _s^2(\varepsilon _s+P_s)v_s^jv_s^k+P_sg^{jk},\label{eq17}
\end{eqnarray}
where $P_s$ is the pressure, and $\varepsilon _s$ is the total energy density defined by
\begin{equation}
\varepsilon _s=m_sn_s+P_s/(\gamma _g-1)\label{eq18}.
\end{equation}
The gas constant $\gamma _g$ take the value  $4/3$ for $T\rightarrow \infty $ and $5/3$ for $T\rightarrow 0$.
The ion temperature profile is closely adiabatic and it approaches $10^{12}\,K$ near the horizon \cite{thirty six}. Far from the (event) horizon electron (positron) temperatures are essentially equal to the ion temperatures, but closer to the horizon the electrons are progressively cooled to about $10^8-10^9\,K$ by mechanisms like multiple Compton scattering and synchrotron radiation. Using the conservation of entropy the equation of state can be expressed by
\begin{equation}
\frac{D}{D\tau }\left(\frac{P_s}{n_s^{\gamma _g}}\right)=0.\label{eq19}
\end{equation}
The full equation of state for a relativistic fluid, as measured in the fluid's rest frame, is as follows \cite{thirty seven,thirty eight}:
\begin{equation}
\varepsilon =m_sn_s+m_sn_s\left[\frac{P_s}{m_sn_s}-\frac{{\rm i}H_2^{(1)^\prime }({\rm i}m_sn_s/P_s)}{{\rm i}H_2^{(1)}({\rm i}m_sn_s/P_s)}\right]\label{eq20},
\end{equation}
where the $H_2^{(1)}(x)$ are Hankel functions. Using Eq. (\ref{eq17}), the energy and momentum conservation equations, Eqs. (\ref{eq14}) and (\ref{eq15}), can be rewritten for each species $s$ in the following form
\begin{align}
&\frac{1}{\alpha }\frac{\partial }{\partial t}P_s-\frac{1}{\alpha }\frac{\partial }{\partial t}[\gamma _s^2(\varepsilon _s+P_s)]-\nabla \cdot [\gamma _s^2(\varepsilon _s+P_s){\bf v}_s]\nonumber\\
&+\gamma _sq_sn_s{\bf E}\cdot {\bf v}_s+2\gamma _s^2(\varepsilon _s+P_s){\bf a}\cdot {\bf v}_s=0,\label{eq21}
\end{align}
and
\begin{align}
&\gamma _s^2(\varepsilon _s+P_s)\left(\frac{1}{\alpha }\frac{\partial }{\partial t}+{\bf v}_s\cdot \nabla \right){\bf v}_s+\nabla P_s\nonumber\\
&-\gamma _sq_sn_s({\bf E}+{\bf v}_s\times {\bf B})+{\bf v}_s\Bigg(\gamma _sq_sn_s{\bf E}\cdot {\bf v}_s\nonumber\\
&+\frac{1}{\alpha }\frac{\partial }{\partial t}P_s\Bigg)+\gamma _s^2(\varepsilon _s+P_s)[{\bf v}_s({\bf v}_s\cdot {\bf a})-{\bf a}]=0\label{eq22}.
\end{align}
The fluid velocities and fields in the above equations are FIDO measured quantities whereas the fluid densities and pressures are measured in the fluid rest frame. Although these equations are valid in a FIDO frame, for $\alpha =1$ they reduce to the corresponding special relativistic equations as given by SK \cite{twenty four} and are valid in a frame in which both the fluids are at rest. Since the plasma is assumed to be falling in radial direction toward the event horizon, the transformation from the FIDO frame to the commoving (fluid) frame involves a boost velocity, which is simply the infall velocity onto the black hole, given by
\begin{equation}
v_{\rm ff}=(1-\alpha ^2)^{\frac{1}{2}},\label{eq23}
\end{equation}
so that, the relativistic Lorentz factor takes the form  $$\gamma _{\rm boost}\equiv (1-v_{\rm ff}^2)^{1/2}=1/\alpha .$$
The linearized two-fluid equations for transverse and longitudinal waves derived in the previous section have the terms $\frac{\partial \alpha }{\partial z}$. Since $\alpha$ is a function of Schwarzschild radial coordinate $r$, it is impossible to solve the two-fluid equations required for these waves analytically, although they form the basis of the numerical procedure. Since the plasma is assumed to be falling in the horizon, the plasma waves varies with the distance from the horizon. The Rindler coordinate system, in which space is locally Cartesian, provides a good approximation to the Schwarzschild metric near the event horizon. In this coordinates the two-fluid equations can be solved in a region very closed to the horizon and cannot be used to the case for extremal black holes because there are no Rindler coordinates locally near horizons. Near the point $(r=2M, \theta=\pi/2,\phi=0)$ the transformation
\begin{eqnarray}
x=2M(\theta -\pi/2), y=2M\phi,z=4M\alpha,\label{eq24}
 \end{eqnarray}
converts the Schwarzschild line element given in Eq. (\ref{eq1}) to
\begin{align}
&ds^2=-\alpha^2dt^2+\left[1-\alpha^2\right]^{-4}dz^2+\left[1-\alpha^2\right]^{-2}(dx^2\nonumber\\
&     +dy^2{\rm cos}^2\frac{x}{2M}),\nonumber\\
&=-\alpha^2dt^2+(dz^2+dx^2+dy^2)\{1+O\left[\alpha^2,(\frac{x}{2M})^2\right]\},\label{eq25}
\end{align}
which is closely approximated by the Rindler geometry with $\alpha=z/4M$ so that
\begin{equation}
\alpha(z)=z\kappa,\hspace{.5cm} \frac{\partial \alpha }{\partial z}=\kappa,\label{eq26}
 \end{equation}
where $\kappa=1/4M$ is the surface gravity of Schwarzschild black hole.

\section{One-dimensional\\ Wave Propagation}\label{sec4}
Since the two-fluid plasma are assumed to be falling toward the horizon due to the strong gravitational field of black hole, the waves can be treated as one-dimensional and propagating in the radial direction. The transverse and longitudinal parts of the two-fluid equations in Schwarzschild coordinates can be separated in a form analogous to that used by SK, i.e., to begin with, the velocities and fields are split into longitudinal and transverse components.
Let $v_{sx}$, $v_{sy}$, and $v_{sy}$ be the velocity components along $x$, $y$, and $z$ direction, we define a new complex transverse fields and velocities by introducing the complex variables
\begin{align}
&v_{sz}(z,t)=u_s(z,t), v_s(z,t)=v_{sx}(z,t)+{\rm i}v_{sy}(z,t),\nonumber\\
&B(z,t)=B_x(z,t)+{\rm i}B_y(z,t),\nonumber\\
&E(z,t)=E_x(z,t)+{\rm i}E_y(z,t)\label{eq27},
\end{align}
and setting
\begin{eqnarray}
v_{sx}B_y-v_{sy}B_x=\frac{\rm i}{2}(v_sB^\ast -v_s^\ast B),\nonumber\\
v_{sx}E_y-v_{sy}E_x=\frac{\rm i}{2}(v_sE^\ast -v_s^\ast E)\label{eq28},
\end{eqnarray}
where the $\ast $ denotes the complex conjugate, we could write the continuity equation, Eq. (\ref{eq8}), in the form
\begin{equation}
\frac{\partial }{\partial t}(\gamma _sn_s)+\frac{\partial }{\partial z}(\alpha \gamma _sn_su_s)=0,\label{eq29}
\end{equation}
and Poisson's equation, Eq. (\ref{eq10}), as
\begin{equation}
\frac{\partial E_z}{\partial z}=4\pi (q_1n_1\gamma _1+q_2n_2\gamma _2).\label{eq30}
\end{equation}
Adding the ${\bf e}_{\hat y}$  component multiplied by $i$ to the ${\bf e}_{\hat x}$ component of the Maxwell\rq s equation, Eq. (\ref{eq11}), the transverse equation for the newly defined transverse fields and velocities takes the form
\begin{eqnarray}
\frac{1}{\alpha }\frac{\partial B}{\partial t}=-{\rm i}\left(\frac{\partial }{\partial z}+\frac{\kappa}{\alpha}\right)E.\label{eq31}
\end{eqnarray}
In similar fashion, we obtain from the Maxwell\rq s equation, Eq. (\ref{eq12}),
\begin{eqnarray}
\frac{1}{\alpha }\frac{\partial E}{\partial t}={\rm i}\left(\frac{\partial }{\partial z}+\frac{\kappa}{\alpha}\right)B-4\pi e (\gamma _2n_2v_2-\gamma _1n_1v_1)\label{eq32}.
\end{eqnarray}
Differentiating Eq. (\ref{eq32}) with respect to $t$ and using Eq. (\ref{eq31}) yields
\begin{align}
&\left(\alpha ^2\frac{\partial ^2}{\partial z^2}+3\alpha \kappa\frac{\partial }{\partial z}-\frac{\partial ^2}{\partial t^2}+\kappa^2\right)E\nonumber\\
&=4\pi e\alpha \frac{\partial }{\partial t}(n_2\gamma _2v_2-n_1\gamma _1v_1)\label{eq33}.
\end{align}
The longitudinal component of the momentum conservation equation, Eq. (\ref{eq22}), can be separated out by split up into its three vector components and the transverse component may then be obtained from the ${\bf e}_{\hat x}$ and ${\bf e}_{\hat y}$ components. The resultant longitudinal and transverse components of the momentum conservation equation are then respectively given by
\begin{align}
&\rho _s\frac{Du_s}{D\tau }=q_sn_s\gamma _s\left(E_z+\frac{{\rm i}}{2}\left(v_sB^{\ast}-v^{\ast}_s B\right)\right)\nonumber\\
&+(1-u^2_s)\rho _sa-u_s\left(q_sn_s\gamma _s{\bf E}\cdot {\bf v}_s+\frac{1}{\alpha }\frac{\partial P_s}{\partial t}\right)-\frac{\partial P_s}{\partial z},\label{eq34}
\end{align}
and
\begin{eqnarray}
\rho _s\frac{Dv_s}{D\tau }=q_sn_s\gamma _s(E-{\rm i}v_sB_z+{\rm i}u_sB)-u_sv_s\rho _sa\nonumber\\
-v_s\left(q_sn_s\gamma _s{\bf E}\cdot {\bf v}_s+\frac{1}{\alpha }\frac{\partial P_s}{\partial t}\right),\label{eq35}
\end{eqnarray}
where ${\bf E}\cdot {\bf v}_s=\frac{1}{2}(Ev_s^\ast +E^\ast v_s)+E_zu_s$ and $\rho _s$ is the total energy density defined by
\begin{equation}
\rho _s=\gamma _s^2(\varepsilon _s+P_s)=\gamma _s^2(m_sn_s+\Gamma _gP_s)\label{eq36}
\end{equation}
with $\Gamma _g=\gamma _g/(\gamma _g-1)$.
In order to investigate the transverse electromagnetic waves it is more convenient to work from a combination of the transverse components of the Maxwell\rq s and momentum conservation equations, Eqs. (\ref{eq31})-(\ref{eq33}) and  (\ref{eq35}). The longitudinal waves can be investigated by combining the longitudinal components of the equation of continuity, Eq. (\ref{eq29}), Poisson equation, Eq. (\ref{eq30}), and the conservation of momentum equation, Eq. (\ref{eq34}).

\section{Linearized Equations for Transverse and Longitudinal Waves}\label{sec5}
We linearize the sets of the two-fluid equations for the transverse and longitudinal waves  by considering a small perturbation. We introduce the quantities as given in BHT of the form
\begin{align}
&u_s(z,t)=u_{0s}(z)+\delta u_s(z,t),\nonumber\\
&n_s(z,t)=n_{0s}(z)+\delta n_s(z,t),\nonumber\\
&v_s(z,t)=\delta v_s(z,t),\hspace{.2cm}P_s(z,t)=P_{0s}(z)+\delta P_s(z,t),\nonumber\\
&\rho _s(z,t)=\rho _{0s}(z)+\delta \rho _s(z,t),\hspace{.2cm}{\bf E}(z,t)=\delta {\bf E}(z,t),\nonumber\\
&{\bf B}_z(z,t)={\bf B}_0(z)+\delta {\bf B}_z(z,t),\hspace{.2cm}{\bf B}(z,t)=\delta {\bf B}(z,t).\label{eq37}
\end{align}
Here, magnetic field has been chosen to lie along the radial ${\bf e}_{\hat z}$ direction. The relativistic Lorentz factor is also linearized such that
$\gamma _s=\gamma _{0s}+\delta \gamma _s,$  where
\begin{eqnarray}
\gamma _{0s}=\left(1-{\bf u}_{0s}^2\right)^{-\frac{1}{2}},\quad \delta \gamma _s=\gamma _{0s}^3{\bf u}_{0s}\cdot \delta {\bf u}_s\label{eq38}.
\end{eqnarray}
Neglecting the product of perturbation terms the conservation of entropy, Eq. (\ref{eq19}), is linearized to
\begin{equation}
\delta P_s=\frac{\gamma _gP_{0s}}{n_{0s}}\delta n_s.\label{eq39}
\end{equation}
Also the total energy density given in Eq. (\ref{eq36}) is linearized to
\begin{equation}
\delta \rho _s=\frac{\rho _{0s}}{n_{0s}}\left(1+\frac{\gamma _{0s}^2\gamma _gP_{0s}}{\rho _{0s}}\right)\delta n_s+2u_{0s}\gamma _{0s}^2\rho _{0s}\delta u_s,\label{eq40}
\end{equation}
where $\rho _{0s}=\gamma _{0s}^2(m_sn_{0s}+\Gamma _gP_{0s})$.\\
The transverse part of the Maxwell\rq s equations, Eqs. (\ref{eq31}) and (\ref{eq32}), and their resultant equation, Eq (\ref{eq33}), are linearized to give
\begin{align}
&\frac{1}{\alpha }\frac{\partial {\delta B}}{\partial t}=-{\rm i}\left(\frac{\partial }{\partial z}+\frac{\kappa}{\alpha}\right){\delta E},\label{eq41}\\
&\frac{1}{\alpha }\frac{\partial {\delta E}}{\partial t}={\rm i} \left(\frac{\partial }{\partial z}+\frac{\kappa}{\alpha}\right){\delta B}-4\pi e (\gamma _2n_2v_2-\gamma _1n_1v_1)\label{eq42},
\end{align}
and
\begin{align}
\left(\alpha ^2\frac{\partial ^2}{\partial z^2}+3\alpha \kappa\frac{\partial }{\partial z}-\frac{\partial ^2}{\partial t^2}+\kappa^2\right)\delta E\nonumber\\
=4\pi e\alpha \left(n_{02}\gamma _{02}\frac{\partial \delta v_2}{\partial t}-n_{01}\gamma _{01}\frac{\partial \delta v_1}{\partial t}\right)\label{eq43}.
\end{align}
Linearizing the transverse part of momentum conservation equation, Eq. (\ref{eq35}), differentiating it with respect to $t$ and than substituting for the magnetic field using Eq. (\ref{eq28}), we obtain
\begin{align}
&\left(\alpha u_{0s}\frac{\partial }{\partial z}+\frac{\partial }{\partial t}-u_{0s}\kappa+\frac{{\rm i}\alpha q_s\gamma _{0s}n_{0s}B_0}{\rho _{0s}}\right)\frac{\partial \delta v_s}{\partial t}\nonumber\\
&-\frac{\alpha q_s\gamma _{0s}n_{0s}}{\rho _{0s}}\left(\alpha u_{0s}\frac{\partial }{\partial z}+\frac{\partial }{\partial t}+u_{0s}\kappa\right)\delta E=0\label{eq44}.
\end{align}
The continuity Eq. (\ref{eq29}), and Poisson's Eq. (\ref{eq30}) are linearized to obtain
\begin{align}
&\gamma _{0s}\left(\frac{\partial }{\partial t}+u_{0s}\alpha \frac{\partial }{\partial z}+u_{0s}\kappa+\gamma _{0s}^2\alpha \frac{du_{0s}}{dz}\right)\delta n_s\nonumber\\
&+\left(\alpha \frac{\partial }{\partial z}+\kappa\right)(n_{0s}\gamma _{0s}u_{0s})+n_{0s}\gamma _{0s}^3\Bigg[u_{0s}\frac{\partial }{\partial t}\nonumber\\
& +\alpha \frac{\partial }{\partial z}+\kappa+\alpha \left(\frac{1}{n_{0s}}\frac{dn_{0s}}{dz}+3\gamma _{0s}^2u_{0s}\frac{du_{0s}}{dz}\right)\Bigg]\delta u_s=0\label{eq45},
\end{align}
and
\begin{align}
&\frac{\partial \delta E_z}{\partial z}=4\pi e(n_{02}\gamma _{02}-n_{01}\gamma _{01})+4\pi e(\gamma _{02}\delta n_2-\gamma _{01}\delta n_1)\nonumber\\
&+4\pi e(n_{02}u_{02}\gamma _{02}^3\delta u_2-n_{01}u_{01}\gamma _{01}^3\delta u_1).\label{eq46}
\end{align}
In similar fashion, the longitudinal part of the momentum conservation equation, Eq. (\ref{eq34}), is linearized to give
\begin{align}
&\left\{\frac{\partial}{\partial t}+u_{0s}\alpha \frac{\partial }{\partial z}+\gamma _{0s}^2\alpha (1+u^2_{0s}) \frac{du_{0s}}{dz}\right\}\delta u_s\nonumber\\
&-\frac{\alpha q_s n_{0s}}{\rho_{0s}\gamma_{0s}}\delta E_z+\left(u_{0s}\alpha \frac{du_{0s}}{dz}+\frac{\alpha}{\rho_{0s}}\frac{dP_{0s}}{dz}+\frac{\kappa}{\gamma_{0s}^2}\right)\nonumber\\
&+\frac{1}{\gamma_{0s}^2 n_{0s}}\Bigg\{\frac{\gamma_{0s}^2 \gamma_g P_{0s}}{\rho_{0s}} \left(u_{0s}\frac{\partial}{\partial t}+\alpha\frac{\partial}{\partial z}\right)\nonumber\\
&+\gamma_{0s}^2 \alpha\frac{\gamma_g P_{0s}}{\rho_{0s}}\left( \frac{1}{P_{0s}}\frac{dP_{0s}}{dz}-\frac{1}{n_{0s}}\frac{dn_{0s}}{dz}\right)\nonumber\\
&+\left(1+\frac{\gamma_{0s}^2 \gamma_g P_{0s}}{\rho_{0s}}\right)\left(u_{0s}\gamma_{0s}^2 \alpha\frac{du_{0s}}{dz}+\kappa\right)\Bigg\}=0\label{eq47}.
\end{align}

\section{Dependence of the Unperturbed Values on $z$}\label{sec6}
Since the plasma is falling toward the event horizon, the unperturbed radial velocity for each fluid species as measured by a FIDO along ${\bf e}_{\hat z}$ is assumed to be the infall velocity given by
\begin{equation}
u_{0s}(z)=v_{\rm ff}(z)=[1-\alpha ^2(z)]^{\frac{1}{2}}\label{eq48}.
\end{equation}
From the continuity equation, Eq. (\ref{eq29}), it follows that
\[
r^2\alpha \gamma _{0s}n_{0s}u_{0s}=\mbox{const.}=r_H^2\alpha _H\gamma _Hn_Hu_H,
\]
where the values with a subscript $H$ are the limiting values at the event horizon. The infall velocity at the horizon becomes unity so that $u_H=1$. Since $u_{0s}=v_{\rm ff}$, $\gamma _{0s}=1/\alpha $; and hence $\alpha \gamma _{0s}=\alpha _H\gamma _H=1$. Also, because $v_{\rm ff}=(r_H/r)^\frac{1}{2}$, the number density for each species can be written as follows:
\begin{equation}
n_{0s}(z)=n_{Hs}v_{\rm ff}^3(z).\label{eq49}
\end{equation}
The equation of state, Eq. (\ref{eq13}), leads to write the unperturbed pressure,
\begin{equation}
P_{0s}(z)=P_{Hs}(\frac{n_{0s}}{n_{Hs}})^{\gamma _g}\label{eq50},
\end{equation}
which in terms of the infall velocity can be written as
\begin{equation}
P_{0s}(z)=P_{Hs}v_{\rm ff}^{3\gamma _g}(z)\label{eq51}.
\end{equation}
Since $P_{0s}=k_Bn_{0s}T_{0s}$, then with $k_B=1$, the temperature profile is
\begin{equation}
T_{0s}=T_{Hs}v_{\rm ff}^{3(\gamma _g-1)}(z)\label{eq52}.
\end{equation}
The unperturbed magnetic field is purely in the radial $z$ direction. It does not experience effects of spatial curvature. From the flux conservation equation $\nabla \cdot {\bf B}_0=0$, it follows that
\[
r^2B_0(r)=\mbox{const.}
\]
from which one obtains the unperturbed magnetic field in terms of the infall velocity to the form
\begin{equation}
B_0(z)=B_Hv_{\rm ff}^4(z).\label{eq53}
\end{equation}
The acceleration of the two-fluid toward the horizon can be determined directly from Eq. (\ref{eq48}) which approximated by the Rindler geometry as
\begin{equation}
\frac{dv_{\rm ff}}{dz}=-\alpha\kappa\frac{1}{v_{\rm ff}}.\label{eq54}
\end{equation}
Similarly, using Rindler geometry the derivatives of the unperturbed fields and fluid quantities with respect to $z$ in terms of infall velocity become
\begin{eqnarray}
\frac{du_{0s}}{dz}&=&-\alpha \kappa\frac{1}{v_{\rm ff}},\qquad \frac{dB_0}{dz}=-4\alpha \kappa\frac{B_0}{v_{\rm ff}^2},\nonumber\\
\frac{dn_{0s}}{dz}&=&-3\alpha \kappa\frac{n_{0s}}{v_{\rm ff}^2},\quad \frac{dP_{0s}}{dz}=-3\alpha \kappa\frac{\gamma _gP_{0s}}{v_{\rm ff}^2}\label{eq55}.
\end{eqnarray}
The equation of continuity, Eq. (\ref{eq45}), and momentum conservation equation, Eq. (\ref{eq47}), with the help of Eq. (\ref{eq55}) in the Rindler coordinates become
\begin{align}
&\gamma _{0s}\left(\frac{\partial }{\partial t}+u_{0s}\alpha \frac{\partial }{\partial z}-\frac{\alpha^2}{u_{0s}}\kappa\right)\delta n_s+\Bigg(\alpha \frac{\partial }{\partial z}+\kappa\Bigg)(n_{0s}\gamma _{0s}u_{0s})\nonumber\\
&+n_{0s}\gamma _{0s}^3\left(u_{0s}\frac{\partial }{\partial t}+\alpha \frac{\partial }{\partial z}+\left(1-\frac{3}{u_{0s}^2}\right)\kappa\right)\delta u_s=0\label{eq56},
\end{align}
and
\begin{align}
\left(\frac{\partial }{\partial t}+u_{0s}\alpha \frac{\partial }{\partial z}-\frac{1+u_{0s}^2}{u_{0s}}\kappa\right)\delta u_s\nonumber\\
+\frac{v^2_{Ts}}{2\gamma^2_{0s}n_{0s}}\left(\alpha \frac{\partial }{\partial t}+u_{0s}\frac{\partial }{\partial z}-\frac{\alpha^2}{u_{0s}^2}\kappa\right)\delta n_s\nonumber\\
-\frac{\alpha q_s n_{0s}}{\rho_{0s}\gamma_{0s}}\delta E_z-\frac{3}{2}\frac{\alpha^2 v^2_{Ts}}{u^2_{0s}}\kappa=0\label{eq57},
\end{align}
where $v^2_{Ts}=2\gamma_g \gamma^2_{0s}P_{0s}/\rho_{0s}$ is the thermal velocity of the fluids of species $s$. Since the $\gamma^2_{0s}$ factor involved in the energy density $\rho_{0s}$ in the denominator, the $\gamma^2_{0s}$ factor in the numerator cancels out and therefore, the thermal velocity of the two-fluid plasmas are frame independent.

\section{The WKB Approximation}\label{sec7}
The work presented in this paper is devoted to the derivation of the local dispersion relations for the transverse and longitudinal waves propagating in two-fluid plasmas falling in radial direction toward the event horizon of a Schwarzschild black hole. The main result of the present work seems to be the discovery of an instability caused by the general relativistic terms in the dispersion relation as shown by Sakai and Kawata \cite{twenty four} for the electron-positron plasma in the special relativistic case. Due to the Einstein\rq s fundamental principle of general relativity, all physical process confined to a small laboratory (i.e., localized processes), we here consider the local values of lapse function $\alpha\approx \alpha_0$ in the range $0<\alpha_0<1$ so that the general relativistic effects can be shown by build up a more complete picture by considering a large number of layers within this range of $\alpha_0 $  values. In this approximation, the lapse function takes $\alpha_0=0$ at the event horizon and $\alpha_0=1$ for special relativity.

Free falling in any gravitational field occur exactly in the same way as they do in a laboratory outside of any gravitational field. Therefore, if no local instability is present in special relativistic plasmas, then no local instability should occur in a frame of reference comoving with a plasma free falling into a black hole. Since the transformation from this comoving frame to a coordinate frame of the metric given in Eq. (\ref{eq1}) is a simple Lorenz boost with the velocity $u_{0s}=v_{ff}$ in the radial direction, no local instability should also appear without considering non-local effect. Perturbations can grow or decay due to non-local effects related to the general relativity as well as to the fact that the initial equilibrium state of the plasma is non-uniform. We assume that the wavelength is small compared with the range over which the equilibrium quantities change significantly. Then the wavelength must be smaller in magnitude in comparison with the scale of the gradient of the lapse function  $\alpha $. The mathematically correct way to include non-local effects is to consider WKB approximation, that is to write all variables as $V = V_a(z) \exp (i \int k(z) dz - i \omega t)$, where $V_a(z)$ and $k(z)$ are functions, which vary slowly on the scale of the wavelength $\lambda = 2\pi/k(z)$, i.e. $\lambda dk/dz \ll 1$ and $\lambda dV_a/dz \ll 1$.

Since the only scale in this work is the radius of the black hole horizon, $r_H$. Taking the Fourier
transformation, the set of two-fluid equations for transverse electromagnetic waves, i.e., Maxwell\rq s Eqs. (\ref{eq41})-(\ref{eq43}) and the transverse part of the momentum conservation Eq. (\ref{eq44}) reduced to give
\begin{align}
&\left(\alpha _0 k(z)+{\rm i}\kappa\right)\delta E+{\rm i}\omega\delta B=0,\label{eq58}\\
&{\rm i}\omega \delta E=\left(\alpha _0 k(z)+{\rm i }\kappa\right)\delta B+4\pi \alpha _0 e(\gamma _{02}n_{02}\delta v_2-\gamma _{01}n_{01}\delta v_1)\label{eq59},
\end{align}
\begin{align}
&\left({\alpha^2_0k(z)^2-{\rm i}3 \alpha _0 \kappa k(z)-\omega ^2-\kappa^2}\right)\delta E\nonumber\\
&={{\rm i}4\pi e\alpha _0\omega (n_{02}\gamma _{02}\delta v_2-n_{01}\gamma _{01}}\delta v_1),\label{eq60}
\end{align}
and
\begin{eqnarray}
\omega \left(\alpha _0k(z)u_{0s}-\omega +{\rm i}u_{0s}\kappa+\frac{\alpha _0q_s\gamma _{0s}n_{0s}B_0}{\rho _{0s}}\right)\delta v_s\nonumber\\
-\frac{{\rm i}\alpha _0 q_s\gamma _{0s}n_{0s}}{\rho _{0s}}\left(\alpha _0k(z)u_{0s}-\omega -{\rm i}u_{0s}\kappa\right)\delta E=0.\label{eq61}
\end{eqnarray}
Similarly, the set of two-fluid equations for longitudinal waves described by Eqs. (\ref{eq46}), (\ref{eq56}), and (\ref{eq57}) when Fourier transformed, become
\begin{eqnarray}
n_{0s}\gamma^2_{0s}\left(\alpha_0 k-u_{0s}\omega+{\rm i}\left(\frac{3}{u^2_{0s}}-1\right)\kappa\right)\delta u_s\nonumber\\
+\left(\alpha_0 u_{0s}k-\omega+\frac{{\rm i}\alpha^2_0}{u_{0s}}\kappa\right)\delta n_s=0,\label{eq62}
\end{eqnarray}
\begin{align}
&\left(\alpha_0 u_{0s}k-\omega+\frac{{\rm i}(1+u_{0s}^2)}{u_{0s}}\kappa\right)\delta u_s+\frac{v^2_{Ts}}{2\gamma^2_{0s}n_{0s}}\Bigg(\alpha_0 k-u_{0s}\omega\nonumber\\
&+\frac{{\rm i}\alpha^2_0}{u_{0s}^2}\kappa\Bigg)\delta n_s+\frac{{\rm i}\alpha_0 q_s n_{0s}}{\rho_{0s}\gamma_{0s}}\delta E_z+\frac{3}{2}\frac{{\rm i}\alpha^2_0 v^2_{Ts}}{u^2_{0s}}\kappa=0\label{eq63},
\end{align}
and
\begin{eqnarray}
{\rm i}k\delta E_z=4\pi e(n_{02}\gamma _{02}-n_{01}\gamma _{01})+4\pi e(\gamma _{02}\delta n_2-\gamma _{01}\delta n_1)\nonumber\\
+4\pi e(n_{02}u_{02}\gamma _{02}^3\delta u_2-n_{01}u_{01}\gamma _{01}^3\delta u_1). \label{eq64}
\end{eqnarray}
Here, the terms arising from the derivatives $dV_a/dz$ and $dk/dz$ in the above two-fluid equations for transverse and longitudinal waves have been neglected. As in any standard WKB approach, the ratio of these neglected terms to the main term is $\sim \lambda/r_H \sim 1/(kr_H) \ll 1$. Since we have already neglected this terms, it is not mathematically correct to keep any other terms of the order of $\sim \lambda/r_H$ in Eqs. (\ref{eq58})-(\ref{eq63}). Since $\kappa\sim 1/(2r_H) \ll 1$, we need to neglect the terms containing $\kappa$ compared to the main term $\alpha_0 k$. Thus, the set of transverse two-fluid equations, Eqs. (\ref{eq58})-(\ref{eq61}) become
\begin{align}
&\alpha _0 k\,\delta E+{\rm i}\omega\delta B=0,\label{eq65}\\
&{\rm i}\omega \delta E=\alpha _0\,k\delta B+4\pi \alpha _0 e(\gamma _{02}n_{02}\delta v_2-\gamma _{01}n_{01}\delta v_1)\label{eq66},\\
&\left({\alpha^2_0k^2-\omega ^2}\right)\delta E={{\rm i}4\pi e\alpha _0\omega (n_{02}\gamma _{02}\delta v_2-n_{01}\gamma _{01}}\delta v_1),\label{eq67}
\end{align}
and
\begin{eqnarray}
\omega \left(\alpha _0ku_{0s}-\omega +\frac{\alpha _0q_s\gamma _{0s}n_{0s}B_0}{\rho _{0s}}\right)\delta v_s\nonumber\\
-{\rm i}\alpha _0\frac{q_s\gamma _{0s}n_{0s}}{\rho _{0s}}\left(\alpha _0ku_{0s}-\omega \right)\delta E=0.\label{eq68}
\end{eqnarray}
Similarly, the longitudinal two-fluid Eqs. (\ref{eq62}) and (\ref{eq63}) follows, as above, to become
\begin{equation}
n_{0s}\gamma^2_{0s}\left(\alpha_0 k-u_{0s}\omega\right)\delta u_s+\left(\alpha_0 u_{0s}k-\omega\right)\delta n_s=0,\label{eq69}
\end{equation}
and
\begin{equation}
\left(\alpha_0 u_{0s}k-\omega\right)\delta u_s
+\frac{v^2_{Ts}}{2\gamma^2_{0s}n_{0s}}\left(\alpha_0 k-u_{0s}\omega\right)\delta n_s+\frac{{\rm i}\alpha_0 q_s n_{0s}}{\rho_{0s}\gamma_{0s}}\delta E_z=0\label{eq70}.
\end{equation}

\section{Dispersion Relations for Transverse \& Longitudinal Waves}\label{sec8}
The dispersion relation describing Alfv\'en and high frequency electromagnetic waves propagating parallel to the unperturbed magnetic field $B_0$ may be obtained by eliminating $\delta v_1$, $\delta v_2$, and $\delta E$ from the transverse parts of Eqs. (\ref{eq67}) and (\ref{eq68}) as
\begin{eqnarray}
k^2-\frac{\omega^2}{\alpha_0^2}=
\frac{\omega_{p1}^2\left(\frac{\omega}{
\alpha_0}-u_{01}k\right)}{u_{01}k-\frac{\omega}{\alpha_0}-\omega_{c1}}
+
\frac{\omega_{p2}^2\left(\frac{\omega}{
\alpha_0}-u_{02}k\right)}{u_{02}k-\frac{\omega}{\alpha_0}+\omega_{c2}},\label{eq71}
\end{eqnarray}
for either the electron-positron or electron-ion plasma. Where the local plasma frequency for transverse wave is $\omega _{ps}=\sqrt{\frac{4\pi e^2\gamma _{0s}^2n_{0s}^2}{\rho _{0s}}}$, which depends on the local number density of electrons and also the local value of the lapse function $\alpha_0$, and the  local cyclotron frequency is $\omega _{cs}=\frac{e^2\gamma _{0s}n_{0s}B_0}{\rho_{0s}}$, which depends upon both the local magnetic field and the lapse function. Note that, the plasma frequency and cyclotron frequency are frame independent (independent of $\gamma_{0s}$) as the thermal velocity of the fluids. This is because a boost to the fluid frame involves the transformation $B_0\rightarrow \gamma_{0s}B_0$ and $\rho _{0s}=\gamma _{0s}^2(m_sn_{0s}+\Gamma _gP_{0s})$. We see that general relativistic effects enter this dispersion relation only in the ratio $\omega/\alpha_0$. If one uses local time $d\tau = \alpha_0 dt$ of the falling observer instead of a global coordinate time $t$, then the dispersion relation is reduced to the special relativistic version as it should be according to the Einstein relativity principle.

Similarly, the dispersion relation for the longitudinal waves can be expressed by eliminating $\delta u_s$, $\delta n_s$, and $\delta E_z$ from Eqs. (\ref{eq64}), (\ref{eq69}), and (\ref{eq70}), of the form
\begin{eqnarray}
1=\Bigg[\frac{\omega^2_{p1}}{(u_{01}k-\frac{\omega}{\alpha_0})^2-\frac{v^2_{T1}}{2}(k-\frac{\omega}{\alpha_0}u_{01})^2}\nonumber\\
+\frac{\omega^2_{p2}}{(u_{02}k-\frac{\omega}{\alpha_0})^2-\frac{v^2_{T2}}{2}(k-\frac{\omega}{\alpha_0}u_{02})^2}\Bigg].\label{eq72}
\end{eqnarray}
Like transverse electromagnetic waves it is also true that the general relativistic effects enter this dispersion relation of longitudinal waves by the ratios of $\omega/\alpha_0$. If we take the local time $d\tau = \alpha_0 dt$ of the FIDO instead of a global time $t$, then the dispersion relation correspond to the special relativistic version as it should be according to the Einstein principle of relativity. It is clear that the dispersion relation for longitudinal waves is independent of local cyclotron frequency $\omega _{cs}$.
If one considers the equivalent case to that of SK \cite{twenty four} for an electron-positron plasma by taking that the two fluids have the same velocity $u_0$, the same equilibrium density $n_0$, and are at the same temperature $T_0$, one obtains from Eq. (\ref{eq72})
\begin{equation}
1=\left[\frac{2\omega^2_{p}}{(u_0k-\frac{\omega}{\alpha_0})^2-\frac{v^2_{T}}{2}(k-\frac{\omega}{\alpha_0}u_{0})^2}
\right].\label{eq73}
\end{equation}
In the limit of zero gravity, i.e., when $\alpha_0\rightarrow 1$ and $u_0\rightarrow 0$, the above Eq. (\ref{eq73}) becomes
\begin{equation}
\omega^2=(2\omega^2_p+k^2 \gamma_gP_0/\rho_0),\label{eq74}
\end{equation}
which is the SK \cite{twenty four} result. The only difference being that $\gamma_g$ has been set to unity in the SK work \cite{twenty four}.
\begin{figure}[h]\label{fig1}
\begin{center}
 \includegraphics[scale=.45]{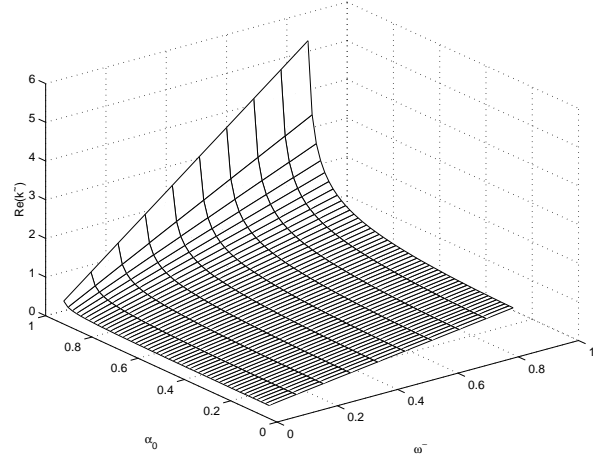}
\end{center}
\caption{\it Purely real Alfv\'en mode for the electron-positron plasma.}
\end{figure}

\begin{figure}[h]\label{fig2}
\begin{center}
 \includegraphics[scale=.45]{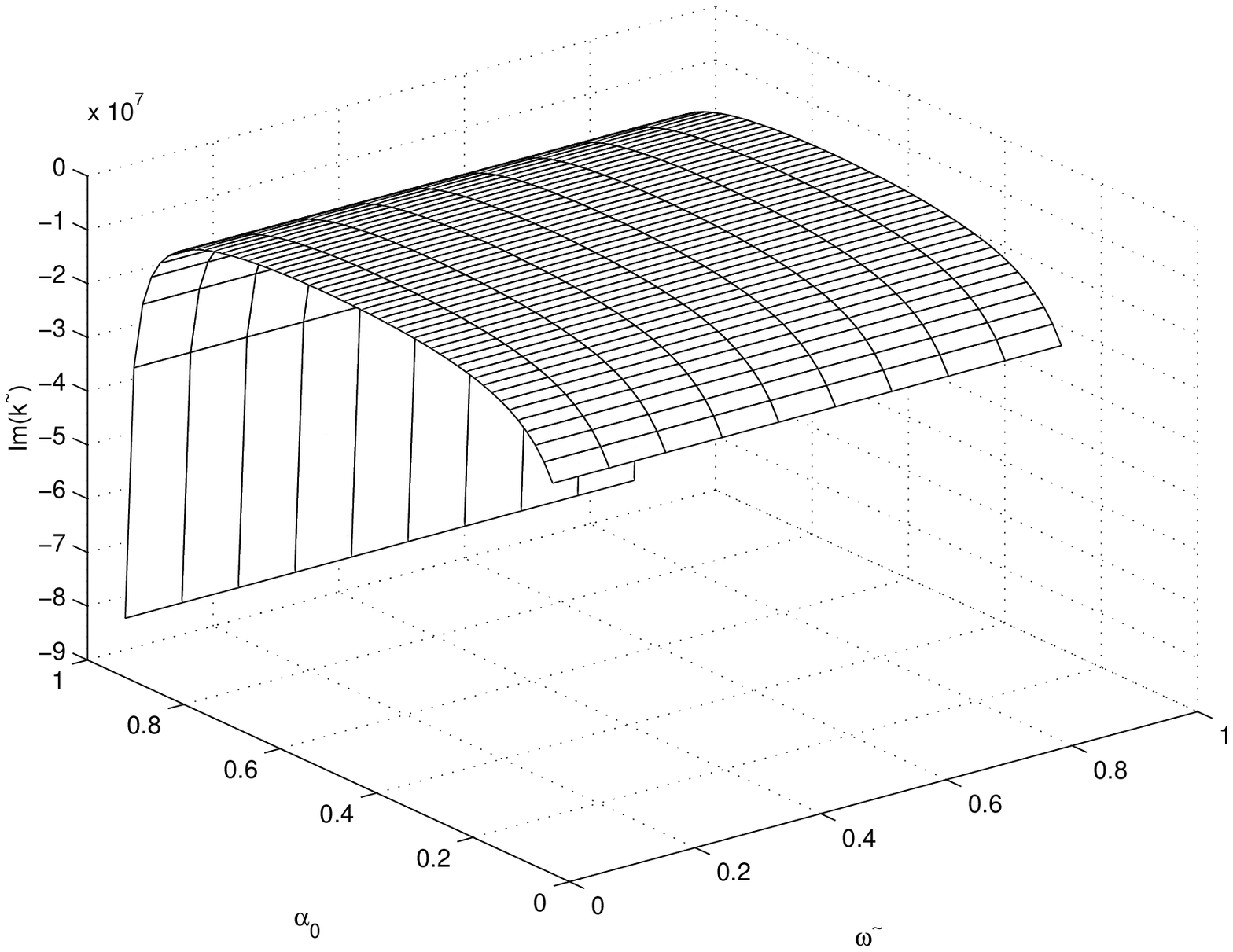}
 \includegraphics[scale=.45]{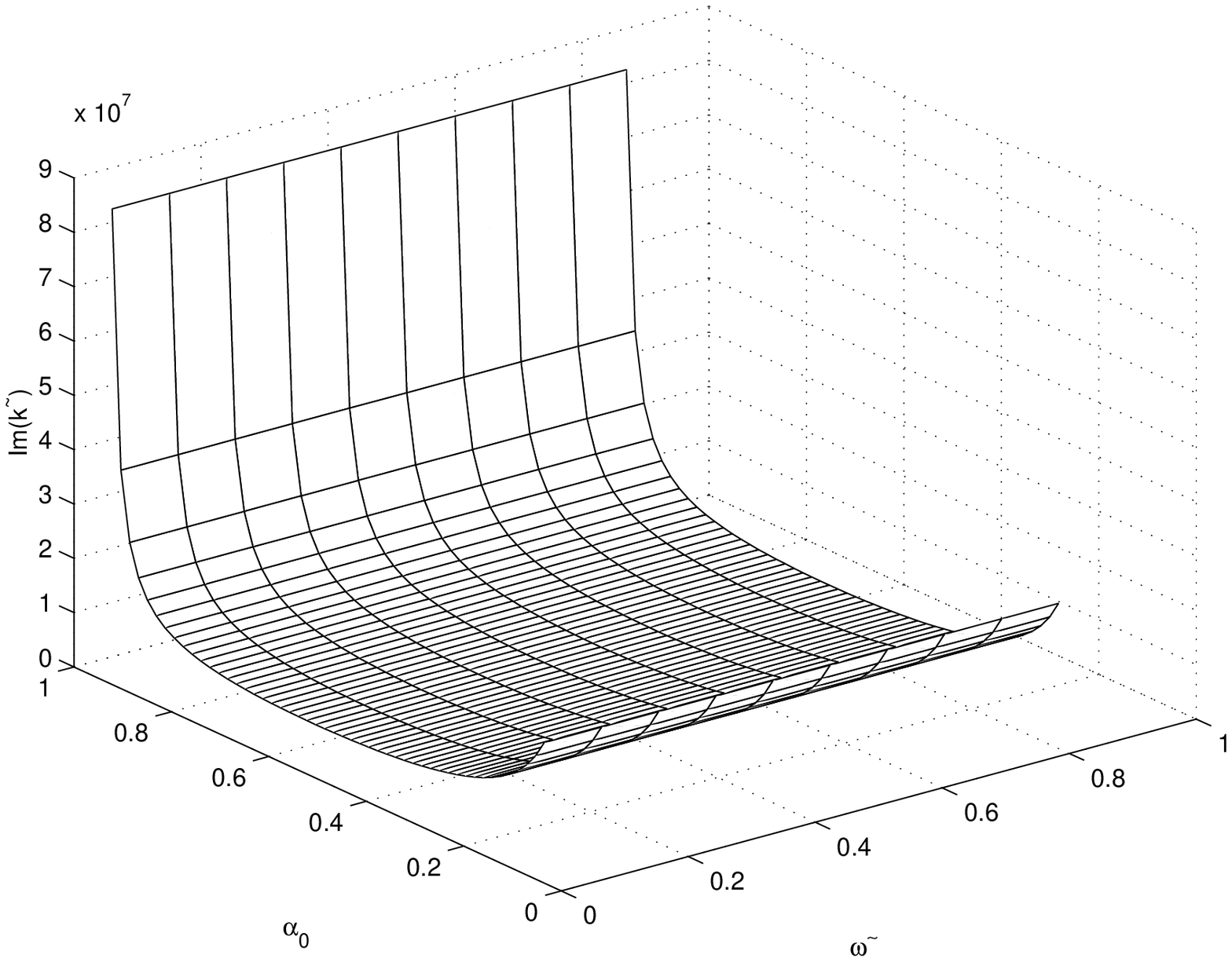}
\end{center}
\caption{\it Left: one purely imaginary Alfv\'en growth mode for the electron-positron plasma.. Right: Another purely imaginary Alfv\'en damped mode.}
\end{figure}
\begin{figure}[h]\label{fig3}
\begin{center}
 \includegraphics[scale=.45]{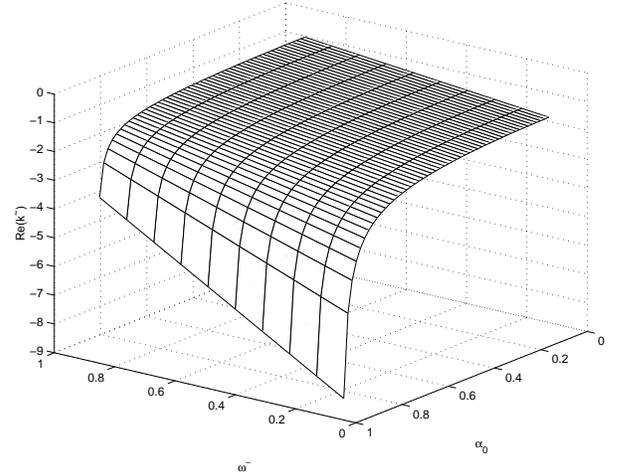}
\caption{\it One purely new real modes for electron-ion plasma. Another is shown in Fig. 1.}
\end{center}
\end{figure}
\section{Numerical Solutions}\label{sec9}
The dispersion relations for the transverse and longitudinal waves given in Eqs. (\ref{eq71}) and (\ref{eq72})  are complicated enough, even in the simplest cases for the electron-positron plasma where both species are assumed to have the same equilibrium parameters, and an analytical solution is cumbersome and unprofitable. We put the sets of two-fluid equations for transverse and longitudinal waves separately in the form of a matrix equation in order to determine all the physically meaningful modes for each waves as follows:
\begin{equation}
(A-kI)X=0.\label{eq75}
\end{equation}
Where the eigenvalue is chosen to be the wave number $k$, the eigenvector $X$ is given by the relevant set of perturbations, and $I$ is the identity matrix. The vector $A$ is the sum of two matrices $A_R$ and $A_I$. The elements of these matrices are, respectively, the real and imaginary terms in the coefficients of the transverse and longitudinal two-fluid equations for perturbation respectively. We need to write the perturbation equations in an appropriate form. We introduce the following set of dimensionless variables:
\begin{align}
&\tilde \omega =\frac{\omega }{\alpha _0\omega _\ast },\quad \tilde k=\frac{kc}{\omega _\ast },\quad \delta \tilde u_s=\frac{\delta u_s}{u_{0s}},\quad \tilde v_s=\frac{\delta v_s}{u_{0s}},\nonumber\\
&\delta \tilde n_s=\frac{\delta n_s}{n_{0s}},\quad \delta \tilde B=\frac{\delta B}{B_0},\quad \tilde E=\frac{\delta E}{B_0},\quad \delta \tilde E_z=\frac{\delta E_z}{B_0}.\label{eq76}
\end{align}
The $\delta u_s$, $\delta v_s$, and $u_{0s}$ are already dimensionless, but it is convenient to defined $\delta \tilde{u_s}$ and $\delta \tilde{v_s}$ as above consistency. This is because the choice of input parameters is the same for each fluid. By considering transverse electromagnetic waves in the gravitational field of Schwarzschild black hole, Daniel and Tajima \cite{twenty three} have shown for electron-positron plasma that the upper branch represents the high frequency electromagnetic waves, which can propagate in vacuum, but cannot exist for frequencies lower then $\sqrt{2\omega _p^2+\omega _c^2}$, which is the cutoff frequency for this plasma and the lower branch represents Alfv\'en waves, which cannot exist for a frequency greater then $\omega _c$, the cyclotron frequency for Alfv\'en waves. Therefore, for electron-positron plasma $\omega _\ast $ has been chosen as
\begin{equation*}
\omega _\ast =\left\{\begin{array}{rl}&\omega _c\hspace{2cm}\mbox{Alfv\'en modes},\\
&\sqrt{2\omega _p^2+\omega _c^2}\qquad \mbox{high frequency modes}\end{array}\right.
\end{equation*}
with $\omega _{p1}=\omega _{p2}$ and $\omega _{c1}=\omega _{c2}$, where $\omega _p=\sqrt{\omega _{p1}\omega _{p2}}$ and $\omega _c=\sqrt{\omega _{c1}\omega _{c2}}$ .
For the case of an electron-ion plasma the choice of $\omega _\ast $  is more complicated matter for transverse waves because both the plasma frequency and the cyclotron frequency are different for each fluid. So it is not clear from the dispersion relation of transverse waves what the natural choice of $\omega _\ast $ should be. It has been assumed for simplicity that
\begin{equation*}
\omega _\ast =\left\{\begin{array}{rl}&\frac{1}{\sqrt{2}}(\omega _{c1}^2+\omega _{c2}^2)^\frac{1}{2}\quad \mbox{Alfv\'en modes},\\
&(\omega _{\ast 1}^2+\omega _{\ast 2}^2)^\frac{1}{2}\qquad \mbox{high frequency modes}\end{array}\right.
\end{equation*}
where $\omega _{\ast s}^2=(2\omega _{ps}^2+\omega _{cs}^2)$. The reason for choosing these values for $\omega _\ast $ is that they reduce to the special relativistic cutoffs in the zero gravity limits for electron-positron plasma. The cutoffs in the special relativistic case are determined by the dispersion relation in that the solutions to the dispersion relation are physical (i.e., $Re(k)>0$) only for certain frequency regimes. As the dispersion relation can not be handled analytically, it is difficult to determine what the cutoffs should be in the case including gravity. Other similar combination for $\omega _\ast $  should not make any real difference to the form of the results because $\omega _\ast $  is really only a scale factor.

The dimensionless eigenvector for the transverse set of equations is
\begin{equation}
\tilde X_{\rm transverse}=\left[\begin{array}{c}\delta \tilde v_1\\\delta \tilde v_2\\\delta \tilde B\\\delta \tilde E\end{array}\right]\label{eq77}.
\end{equation}
Using Eq. (\ref{eq76}), the set of transverse two-fluid equations, Eqs. (\ref{eq65}), (\ref{eq66}), and (\ref{eq68}), become the following dimensionless form:
\begin{eqnarray}
\tilde k\delta \tilde v_s=\left(\frac{\tilde \omega }{u_{0s}}-\left(\frac{q_s}{e}\right)\frac{\omega _{cs}}{u_{0s}\omega _\ast }\right)\delta \tilde v_s+\left(\frac{q_s}{e}\right)\frac{\omega _{cs}}{u_{0s}\omega _\ast }\delta \tilde B\nonumber\\
-{\rm i}\left(\frac{q_s}{e}\right)\frac{\omega _{cs}}{u_{0s}\omega _\ast }\delta \tilde E,\label{eq78}
\end{eqnarray}
\begin{equation}
\tilde k\delta \tilde E=-{\rm i}\tilde \omega \delta \tilde B,\label{eq79}
\end{equation}
and
\begin{equation}
\tilde k\delta \tilde B=u_{01}\frac{\omega _{p1}^2}{\omega _{c1}\omega _\ast }\delta \tilde v_1-u_{02}\frac{\omega _{p2}^2}{\omega _{c2}\omega _\ast }\delta \tilde v_2+{\rm i}\tilde \omega \delta \tilde E\label{eq80}.
\end{equation}
These equations are now in the required form to be used as input to Eq. (\ref{eq75}) for transverse electromagnetic waves.
\begin{figure}[h]\label{fig4}
\begin{center}
 \includegraphics[scale=.40]{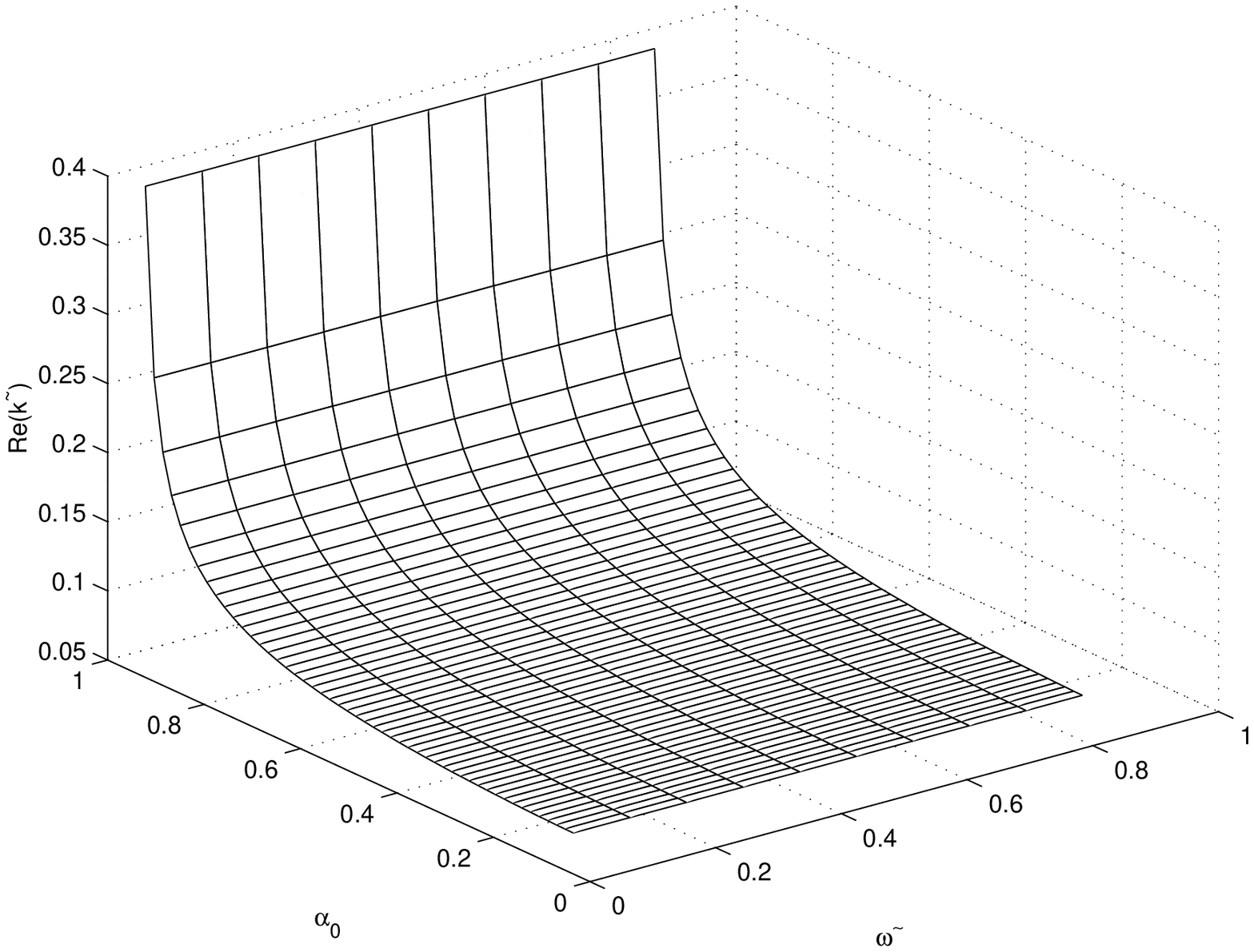}\\
 \includegraphics[scale=.40]{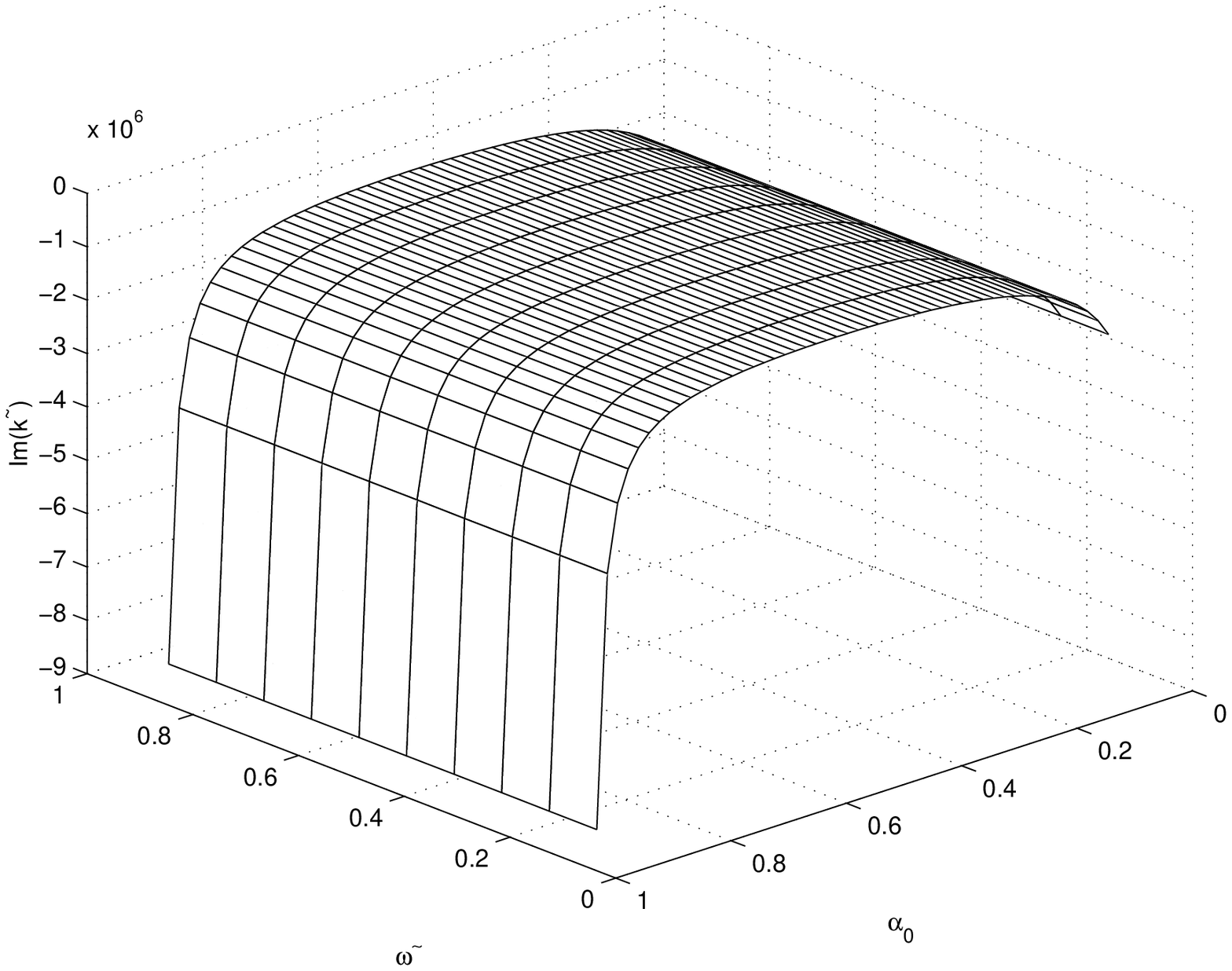}
 \includegraphics[scale=.40]{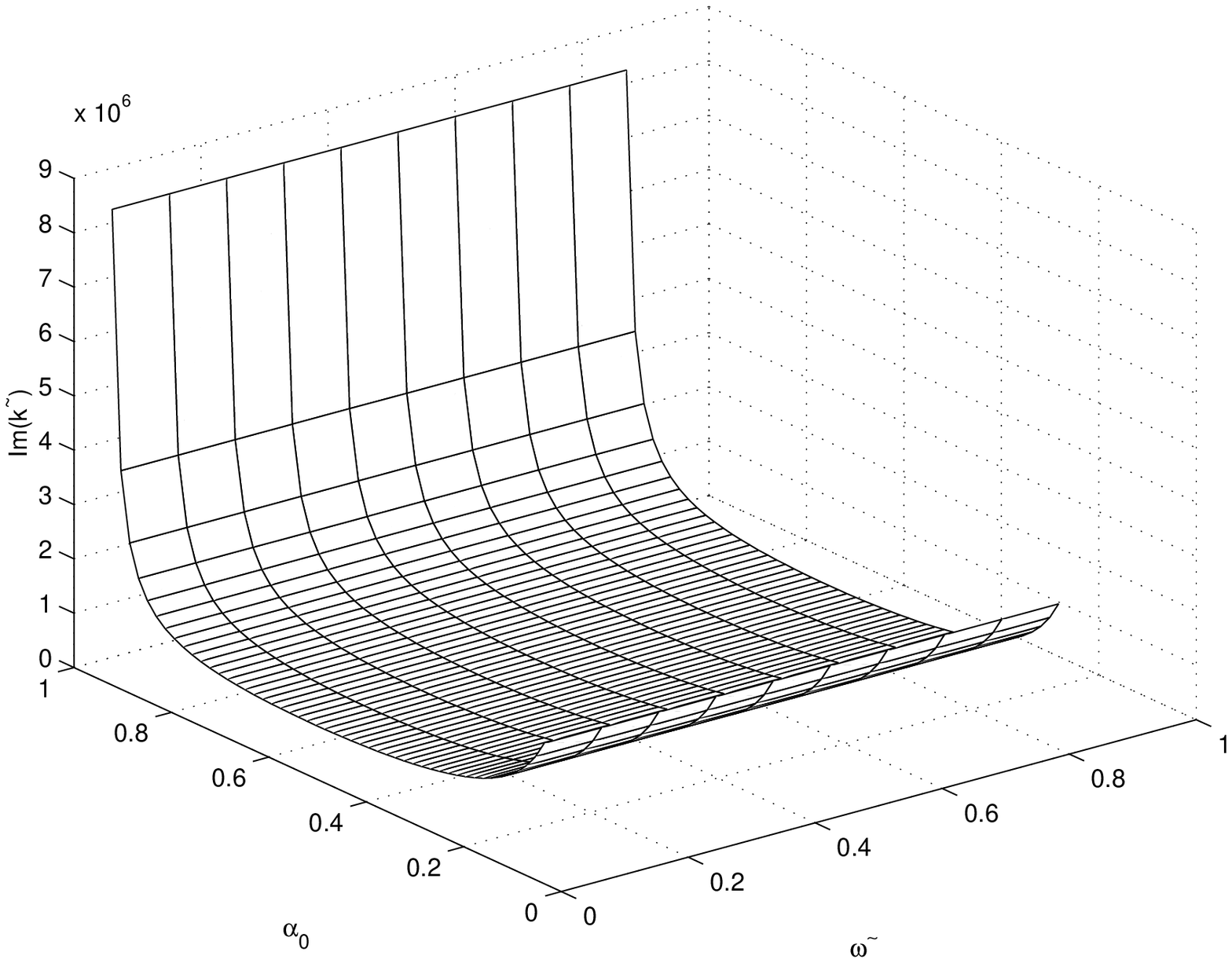}
\end{center}
\caption{\it {\bf Top}: Real part of the complex conjugate pair of Alfv\'en damping and growth modes for the electron-ion plasma. {\bf Bottom}: Left: Imaginary part of the growth mode. Right: Imaginary part of the damped mode. }
\end{figure}
For the longitudinal waves $\omega _\ast $ and  the dimensionless eigenvector have been taken as
\begin{equation*}
\omega _\ast =\left\{\begin{array}{rl}&\omega _p\hspace{1.6cm}\mbox{electron-positron plasma},\\
&\sqrt{\omega _{p1}\omega _{p2}}\qquad \mbox{electron-ion plasma}\end{array}\right.
\end{equation*}
and
\begin{equation}
\tilde X_{\rm Longitudinal}=\left[\begin{array}{c}\delta \tilde u_1\\\delta \tilde u_2\\\delta \tilde n_1\\\delta \tilde n_2\\\delta \tilde E_z\end{array}\right]\label{eq81}.
\end{equation}
\begin{figure}[h]\label{fig5}
\begin{center}
 \includegraphics[scale=.43]{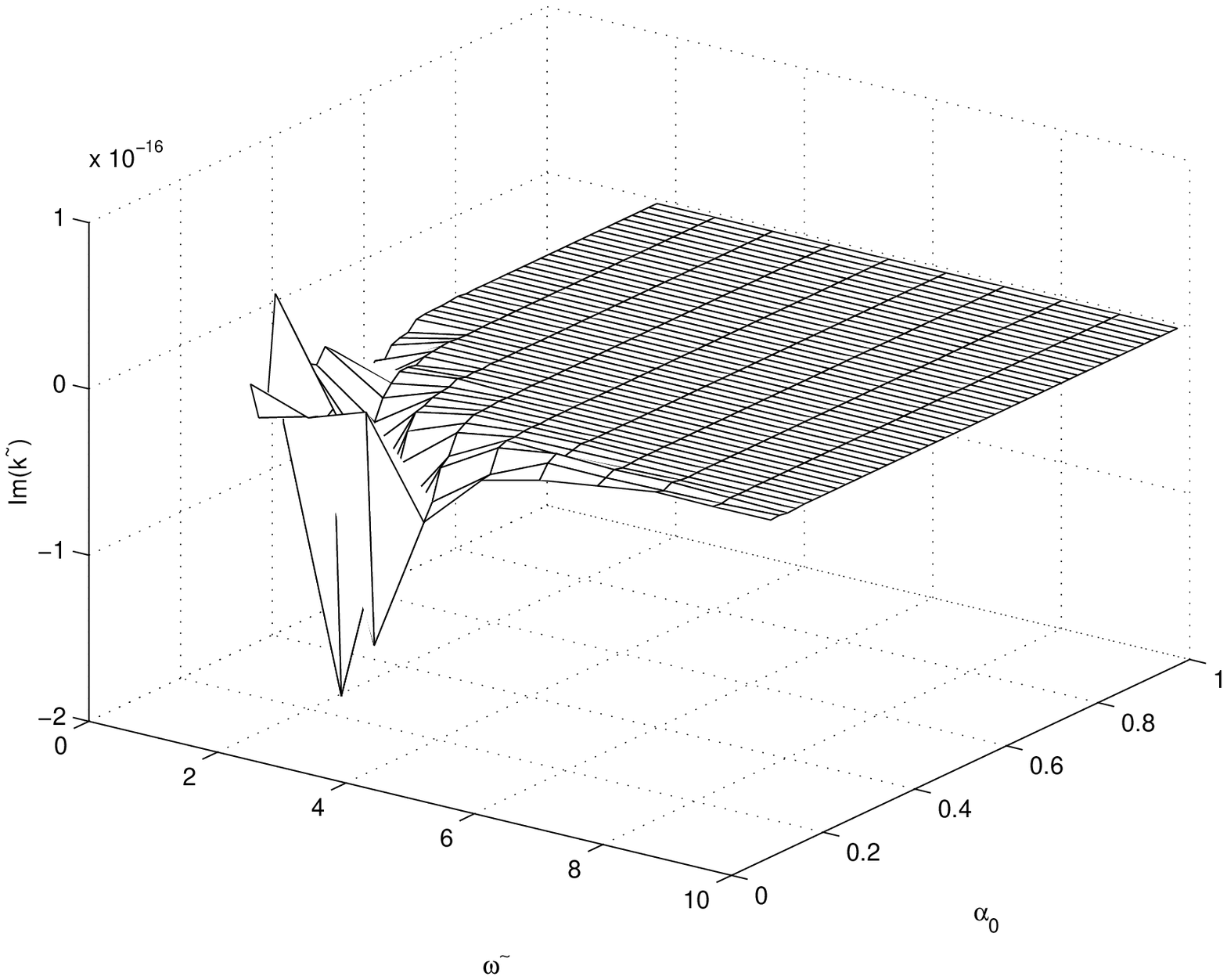}
 \includegraphics[scale=.43]{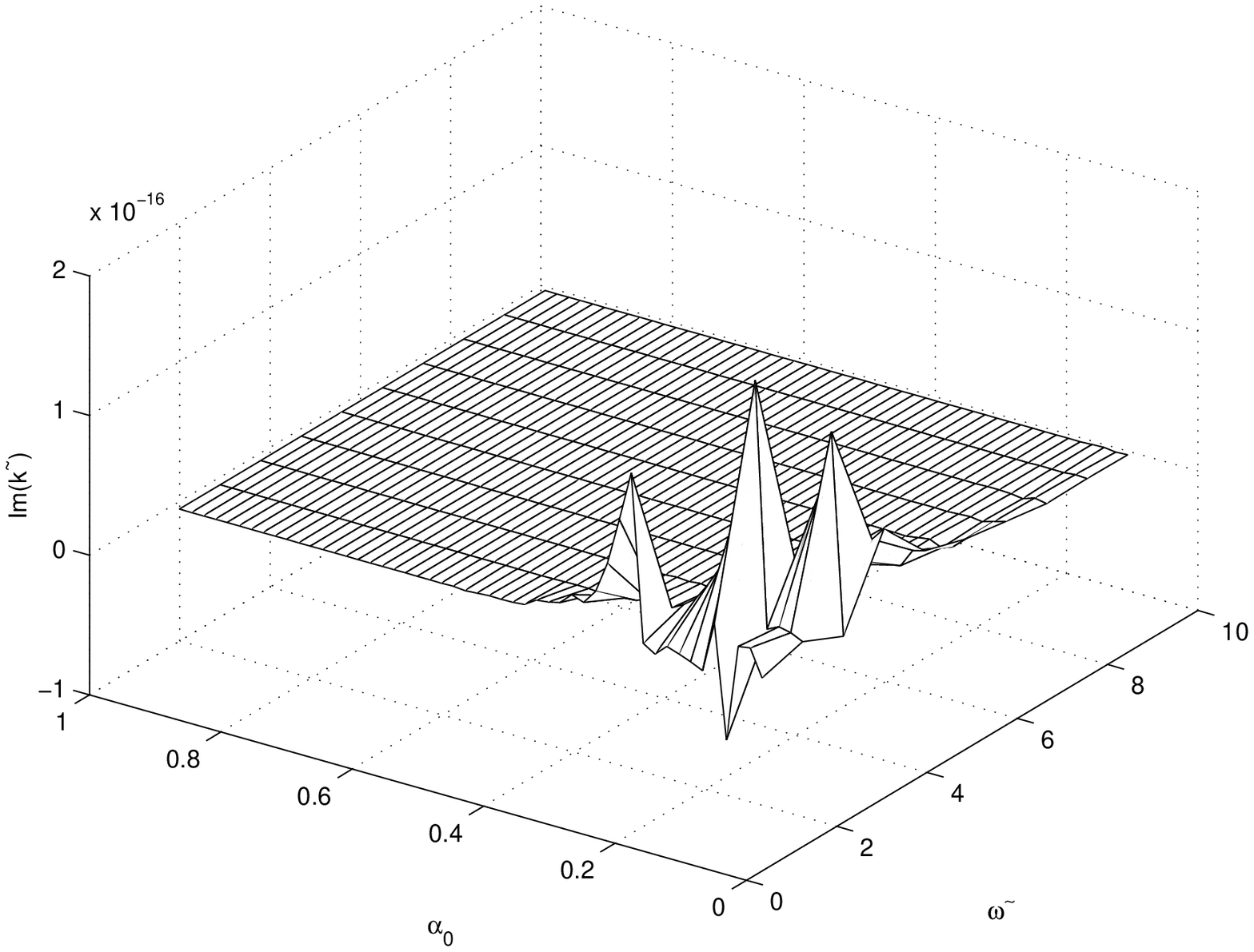}
\end{center}
\caption{\it The left and right imaginary parts show damped and growing occur closed to the horizon. The real part of these modes is shown in Fig. 1.}
\end{figure}
The dimensionless set of equations for longitudinal waves, Eqs. (\ref{eq64}), (\ref{eq69}), and (\ref{eq70}), can be obtained using Eq. (\ref{eq76}) as
\begin{align}
&\tilde k\delta\tilde u_s=\frac{u_{0s}\tilde\omega}{u^2_{0s}-v^2_{Ts}/2}\left(1-\frac{v^2_{Ts}}{2}\right)\delta \tilde u_s\nonumber\\
&-\frac{1}{\gamma^2_{0s}}\frac{1}{u^2_{0s}-v^2_{Ts}/2}\frac{v^2_{Ts}\tilde\omega}{2u_{0s}\gamma^2_{0s}}\delta\tilde n_s-\frac{{\rm i}(q_s/e)\omega_{cs}}{\gamma^2_{0s}\omega_p({u^2_{0s}-v^2_{Ts}/2})}\delta\tilde E_z, \label{eq82}
\end{align}
\begin{align}
&\tilde k\delta\tilde u_s=\frac{u_{0s}\tilde\omega}{u^2_{0s}-v^2_{Ts}/2}\left(1-\frac{v^2_{Ts}}{2}\right)\delta \tilde n_s-\frac{u_{0s}\tilde\omega}{u^2_{0s}-v^2_{Ts}/2}\delta\tilde u_s\nonumber\\
&+\frac{{\rm i}(q_s/e)\omega_{cs}}{\omega_p({u^2_{0s}-v^2_{Ts}/2})}\delta\tilde E_z, \label{eq83}
\end{align}
and
\begin{align}
&\tilde k\delta \tilde E_z={\rm i}u^2_{01}\gamma^2_{01}\frac{\omega _{p1}^2}{\omega _{c1}\omega _p }\delta \tilde u_1-{\rm i}u^2_{02}\gamma^2_{02}\frac{\omega _{p2}^2}{\omega _{c2}\omega _p }\delta \tilde u_2\nonumber\\
&+{\rm i}\frac{\omega _{p1}^2}{\omega _{c1}\omega _p }\delta \tilde n_1-{\rm i}\frac{\omega _{p2}^2}{\omega _{c2}\omega _p }\delta \tilde n_2\label{eq84}.
\end{align}
From the above Eqs. (\ref{eq82}) and (\ref{eq83}) it is clear that there occurs a singularity at the point for which the infall velocity equals to the half of the fluid thermal velocity, called the transonic radius occur at $\alpha_t$ so that $\alpha_t=u^2_{0s}=\frac{v^2_{T_s}}{2}$, for each fluid of species $s$.
\begin{figure}[h]\label{fig6}
\begin{center}
 \includegraphics[scale=.43]{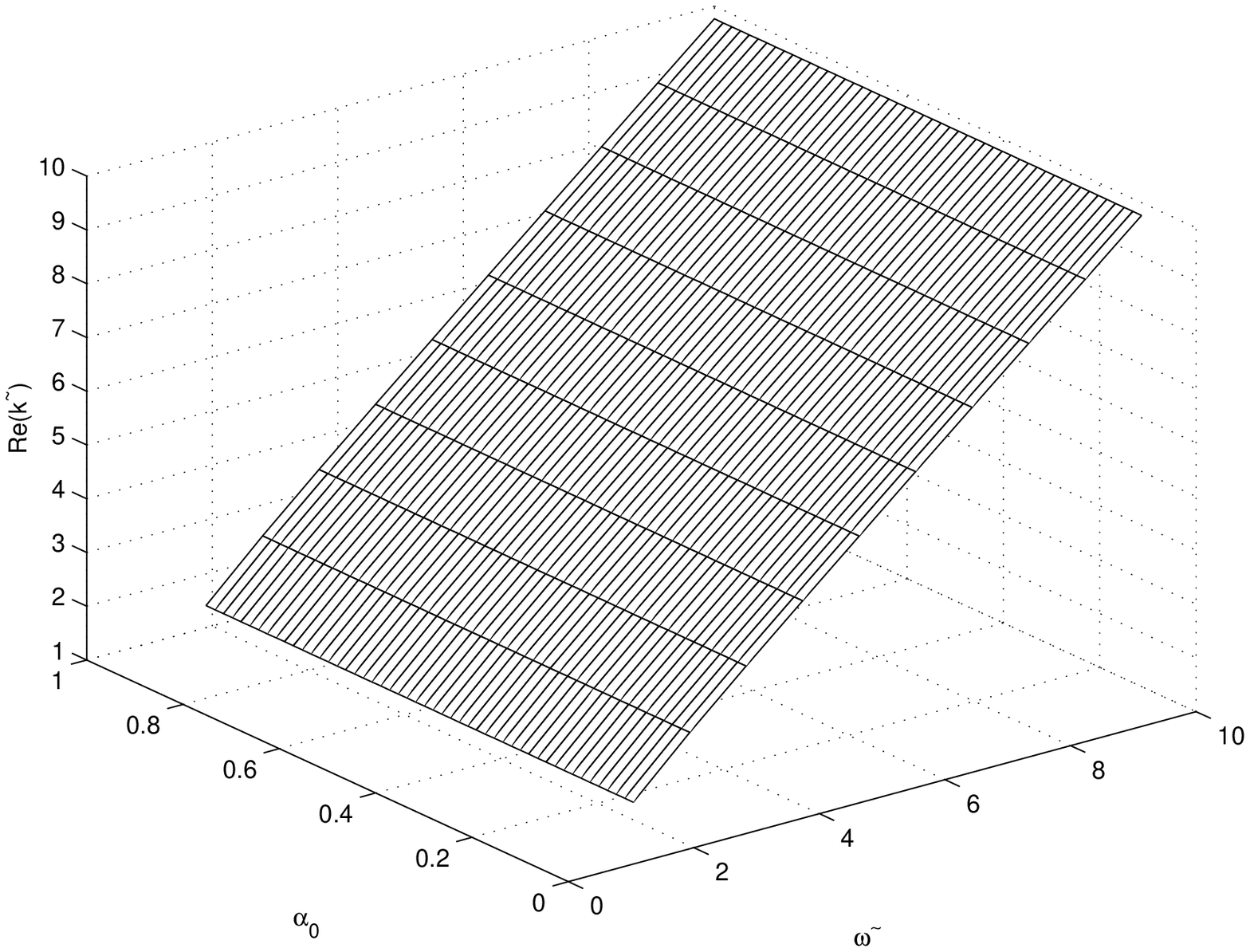}
 \includegraphics[scale=.43]{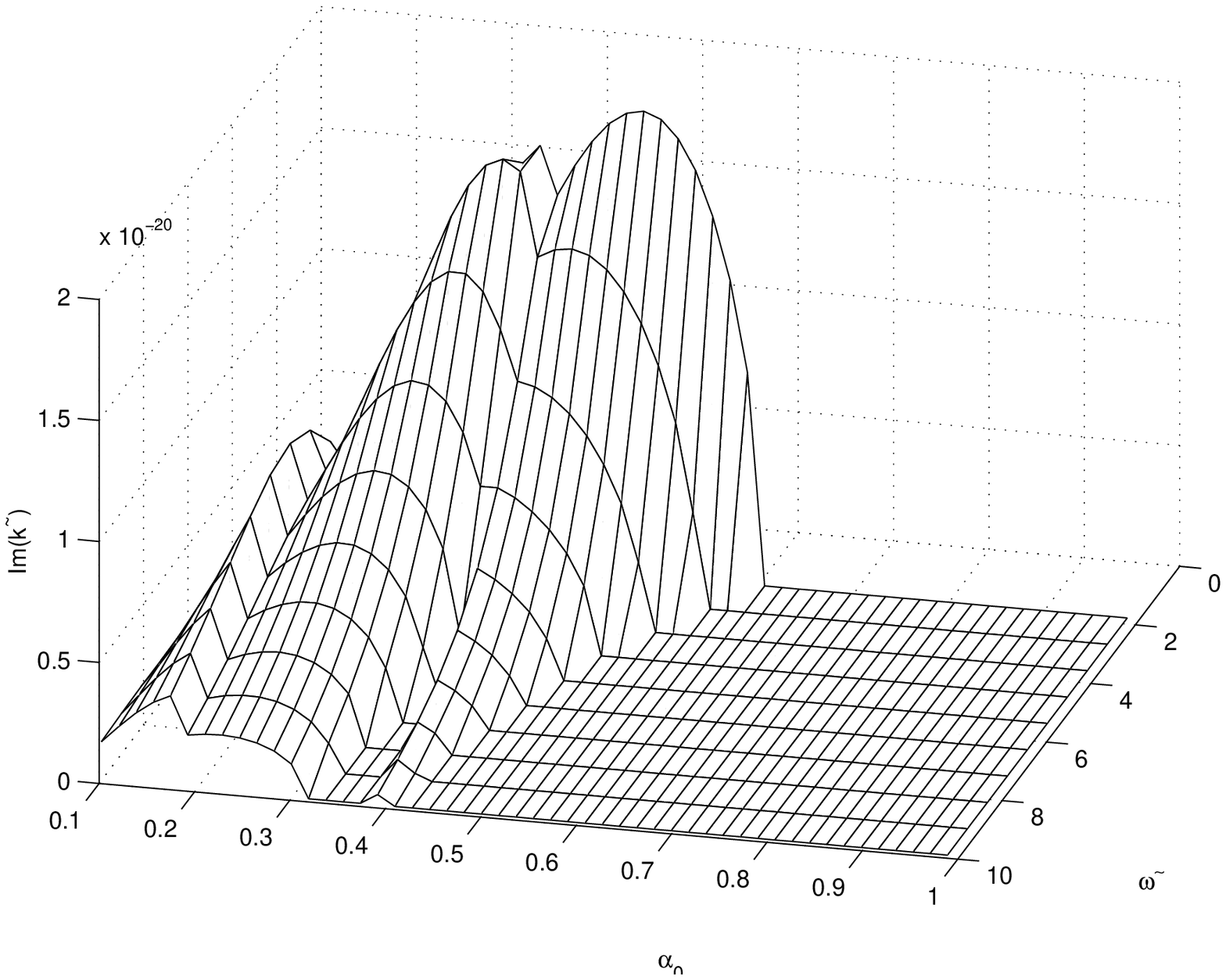}
\end{center}
\caption{\it The left mode is the real part of high frequency  mode for the electron-positron plasma. The right mode show the plasma is not uniform near the horizon and become damped and growing.}
\end{figure}
The position of the transonic radius for each fluid is principally depend on the fluid temperature, the limiting temperature of each fluid at the horizon,  which determines the temperature at any given radius. Therefore, the transonic radius plays a significant role for the longitudinal waves and its influence on the on the waves will be clearly shown by some of the longitudinal waves modes. Equations (\ref{eq82}), (\ref{eq83}), and (\ref{eq84}) are the required equations to be used as input to Eq. (\ref{eq75}) for longitudinal waves which can be put as $(\tilde A-\tilde kI)\tilde X=0$.  The eigenvalues $\tilde k$ of the complex matrix $\tilde A$ have been calculated to draw the modes for the transverse electromagnetic and longitudinal waves using MATLAB.
\begin{figure}[h]\label{fig7}
\begin{center}
 \includegraphics[scale=.4]{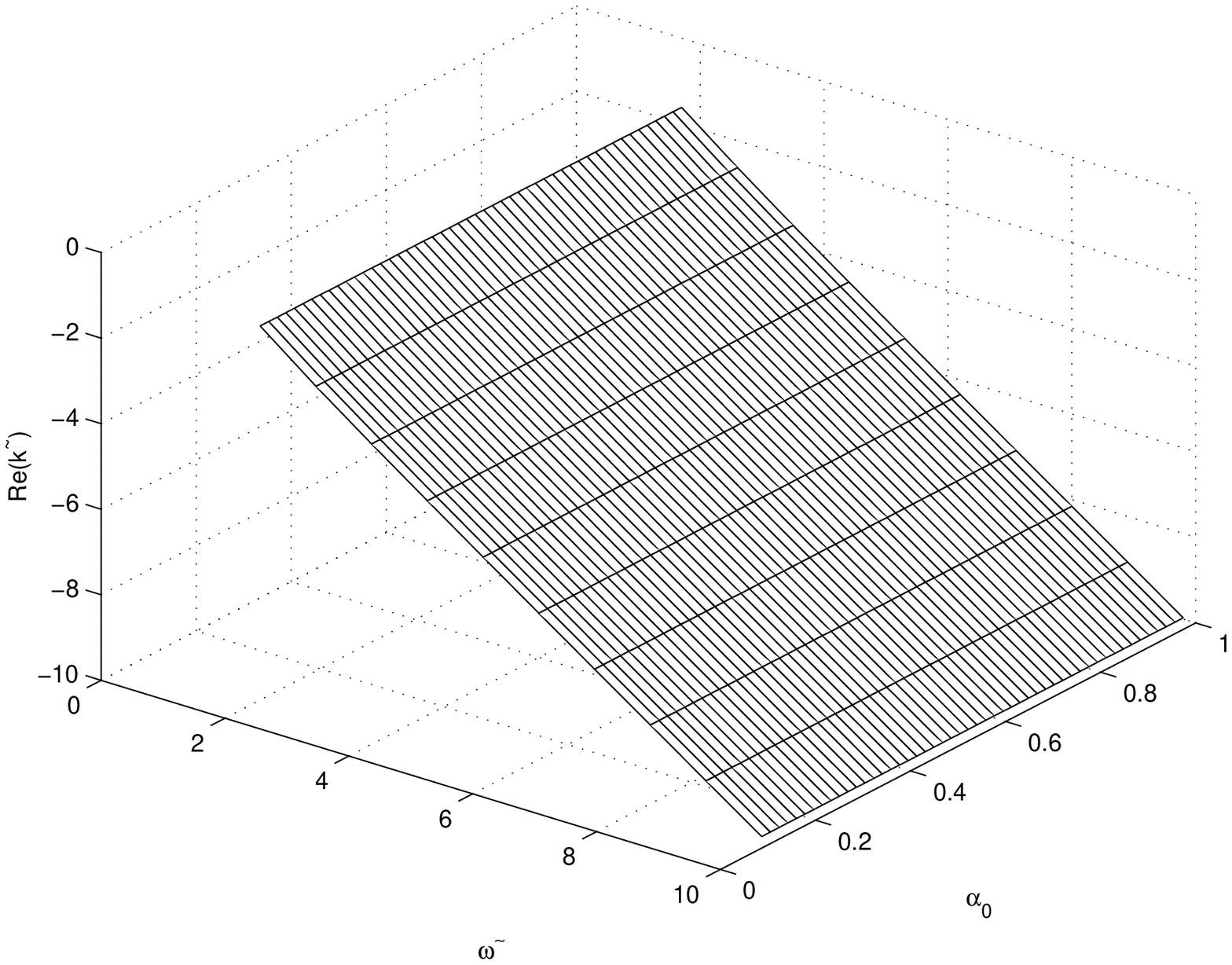}
 \includegraphics[scale=.4]{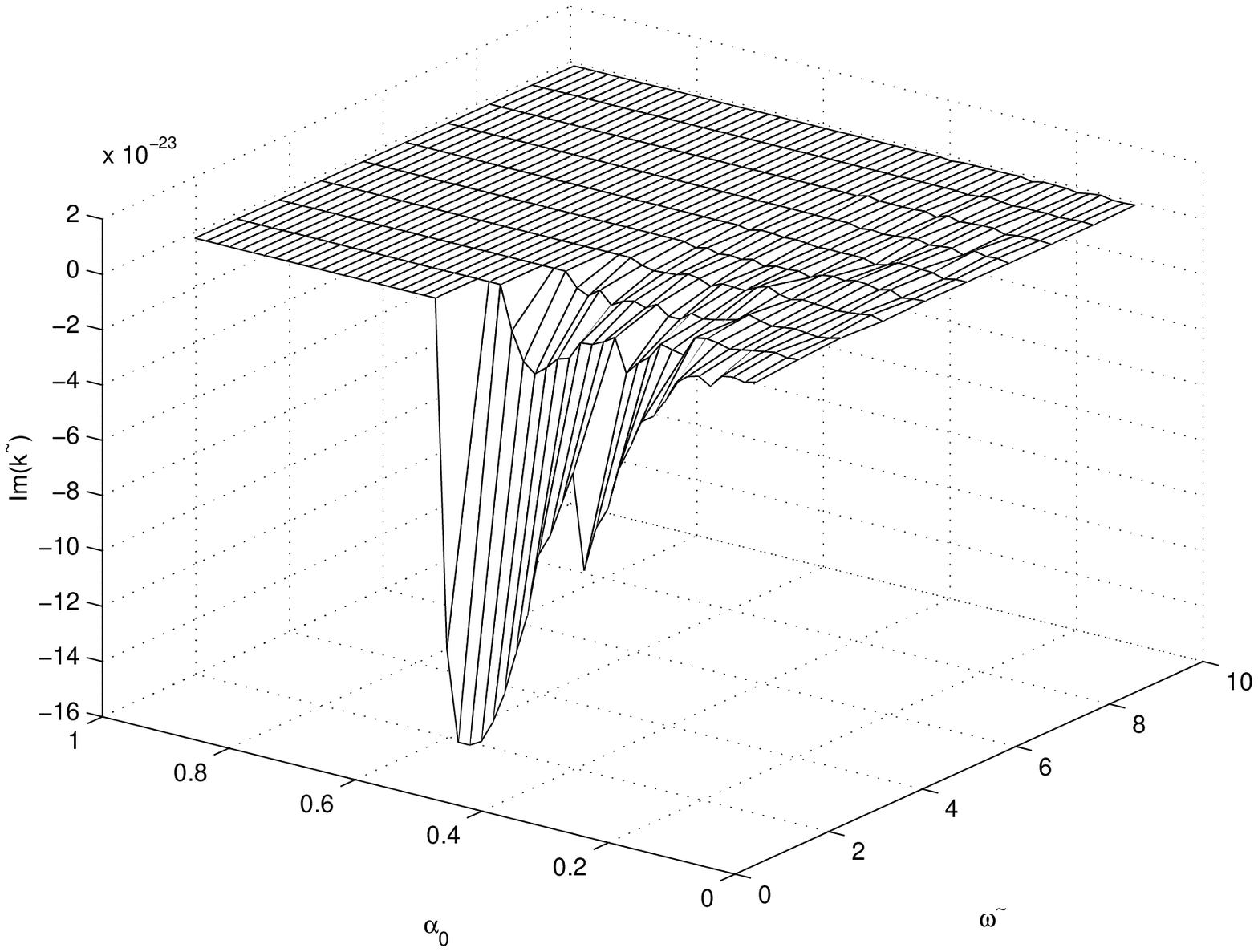}
\end{center}
\caption{\it The left mode is the real part of the two complex conjugate high frequency  mode for the electron-positron plasma. The right mode show the plasma is not uniform near the horizon and become growth near the horizon.}
\end{figure}
\begin{figure}[h]\label{fig8}
\begin{center}
 \includegraphics[scale=.43]{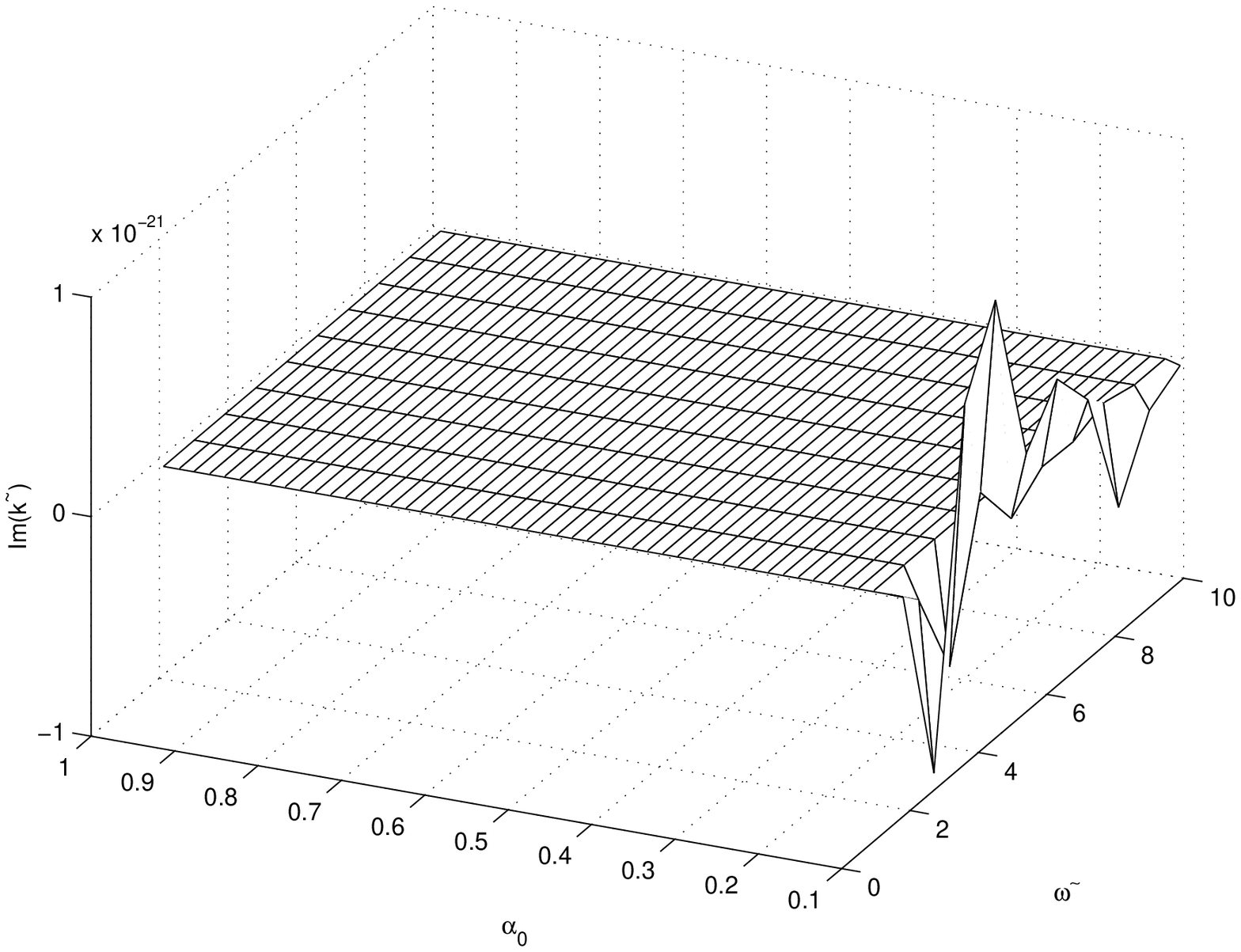}
 \includegraphics[scale=.43]{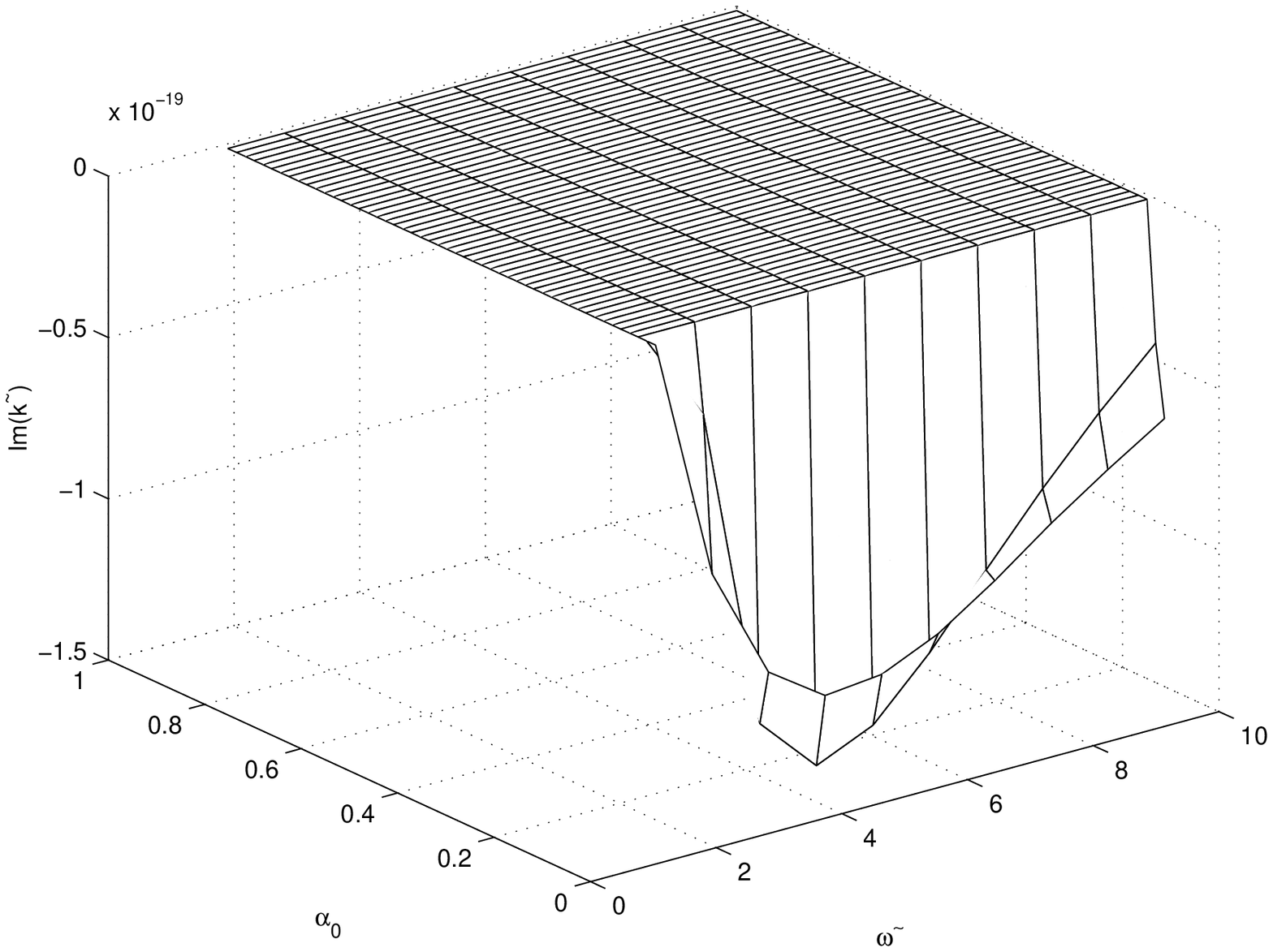}
\end{center}
\caption{\it Both the modes are stable for all the values of $\alpha_0$ but very closed to the horizon they are unstable and the damped and growing occur for all the frequencies. The real part is shown in Fig. 6.}
\end{figure}
\subsection{Results}\label{sec9.1}
We have considered both the electron-positron plasma and the electron-ion plasma. The limiting horizon values for the electron-positron plasma are taken to be
\begin{align*}
n_{Hs}=10^{18}{\rm cm}^{-3},\;T_{Hs}=10^{10}{\rm K},
B_H=3\times 10^6{\rm G},\;\mbox{and}\;\gamma _g=\frac{4}{3}.
\end{align*}
For the electron-ion plasma, the ions are essentially nonrelativistic, and the limiting horizon values are chosen to be
\begin{align*}
n_{H1}=10^{18}{\rm cm}^{-3},\;T_{H1}=10^{10}{\rm K},
n_{H2}=10^{15}{\rm cm}^{-3},\\
\mbox{and}\;T_{H2}=10^{12}{\rm K}.
\end{align*}
The equilibrium magnetic  field has the same value as it has for the electron-positron case. The limiting horizon temperature for each species has been chosen as derived by Colpi et al. \cite{thirty six} from studies of two temperature models of spherical accretion onto black holes. Those values of the limiting horizon densities and the limiting horizon field are arbitrarily chosen which appear to be not inconsistent with current ideas. The gas constant has been chosen as $\gamma _g=\frac{4}{3}$.

\subsection{Alfv\'en Modes}\label{subsec8.2}
\subsubsection{Electron-positron Plasma}\label{subsubsec8.2.1}
For the special relativistic electron-positron plasma, only one purely real Alfv\'en  mode was found to exist by SK \cite{twenty four}. Because both the left and right circularly polarized modes were described by the same dispersion relation. Our work presented in this paper also shows one purely real Alfv\'en mode with two new purely imaginary conjugate modes to exist for the same plasma due to the gravitational field of Schwarzschild black hole. The real mode shown in Fig. 1 is a common mode for special and general relativistic electron-positron plasma but the two purely imaginary modes shown in Fig. 2 are respectively damped and growing and the damped and growing rate are clearly frequency independent, depend only on lapse function $\alpha_0$. The damped mode demonstrates the energy drain from the waves by the gravitational field and the growth mode indicates that the gravitational field is, in fact, feeding energy into the waves. Both the modes are stable for all the frequencies and at all radial distances from the event horizon through $\alpha _0$. Since we are using the convention $e^{{\rm i}\int k(z)dz}=e^{{\rm i}\int[{\rm Re}(k)+{\rm iIm}(k)]dz}$, the damping corresponds to $ {\rm  Im({k})}>0 $ and growth to $\rm  Im(k)<0 $.
\begin{figure}[h]\label{fig9}
\begin{center}
 \includegraphics[scale=.43]{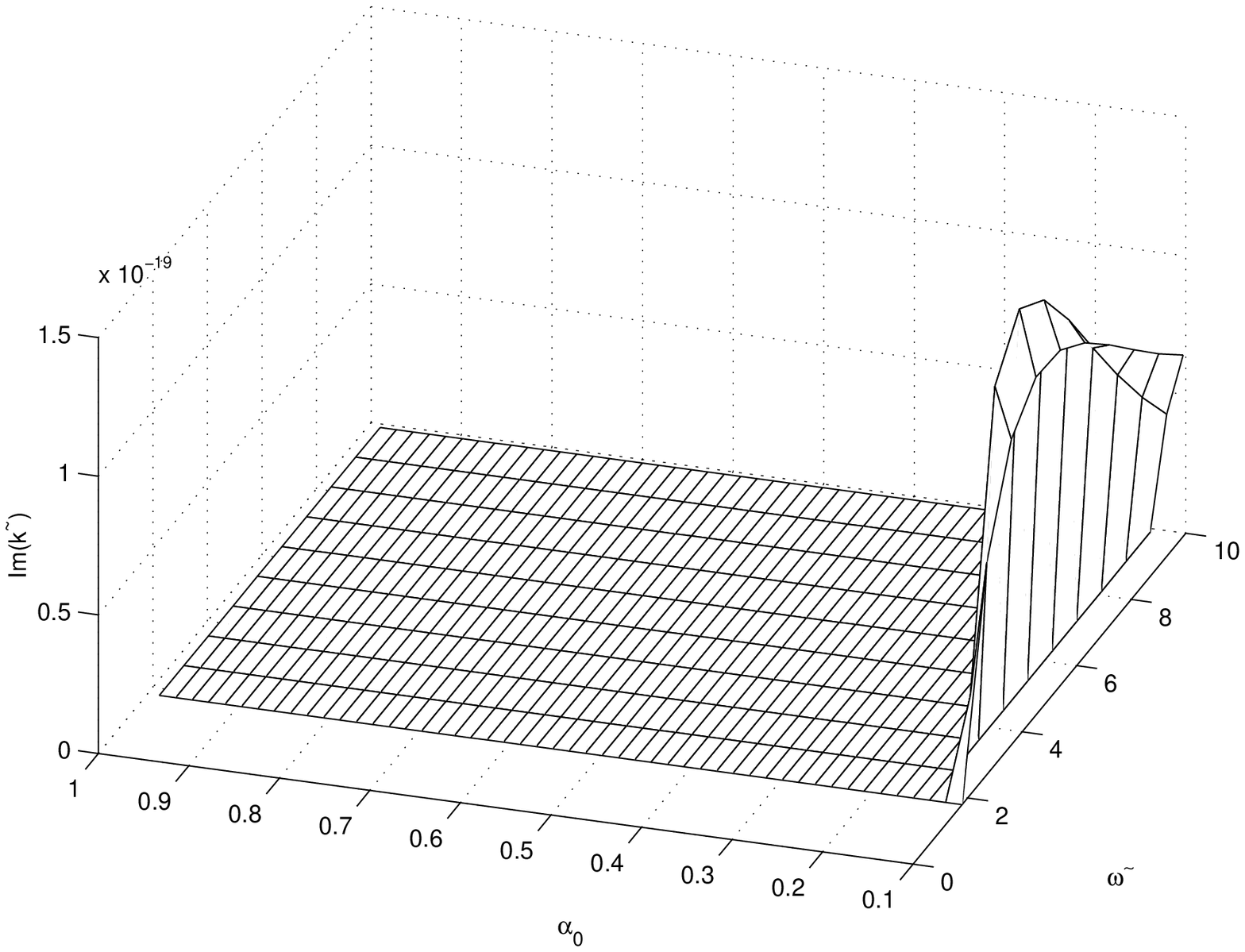}
 \includegraphics[scale=.43]{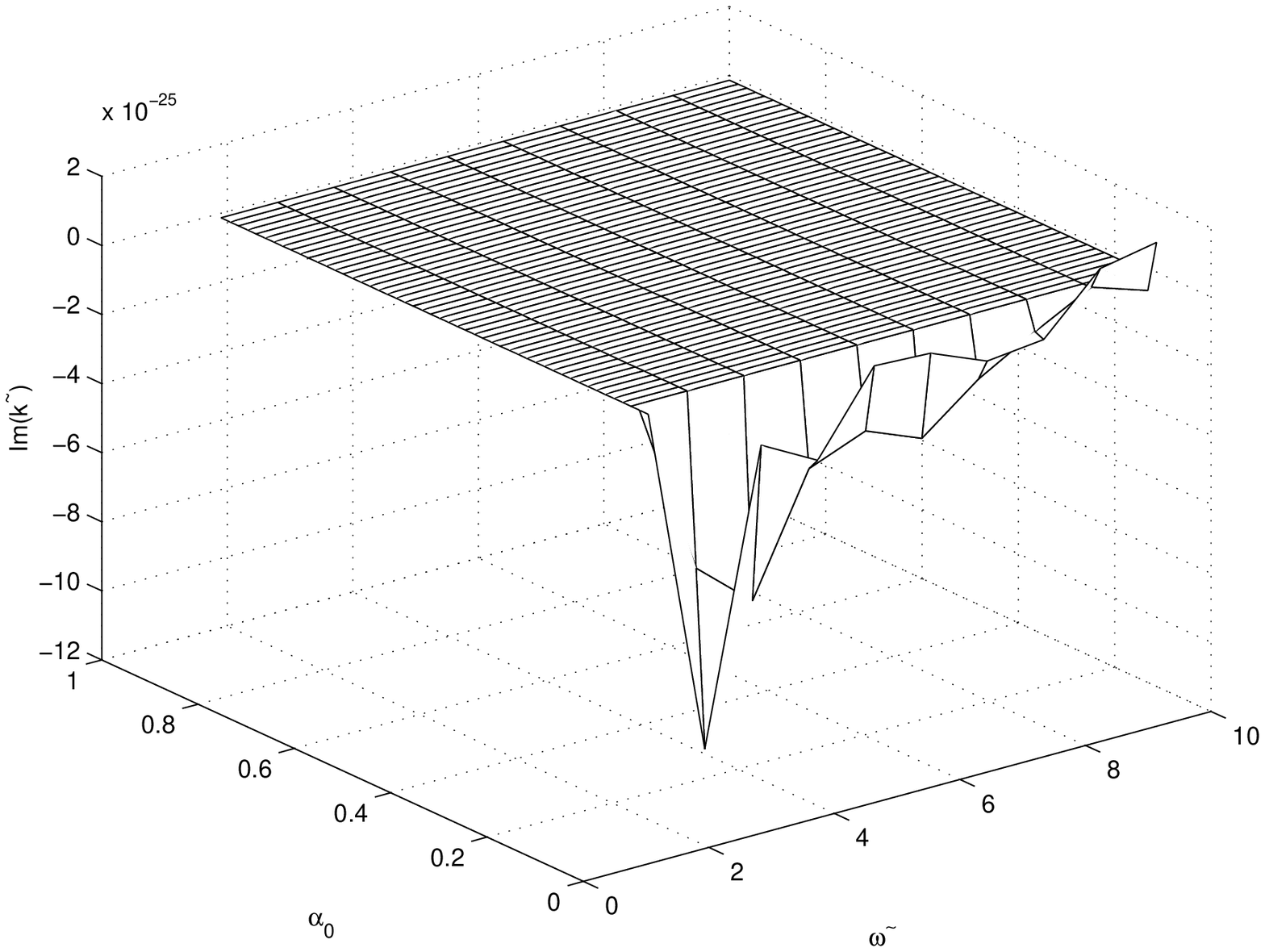}
\end{center}
\caption{\it Instability occur very closed to the horizon for electron-ion plasma. The real part is shown in Fig. 7.}
\end{figure}
\begin{figure}[h]\label{fig10}
\begin{center}
 \includegraphics[scale=.4]{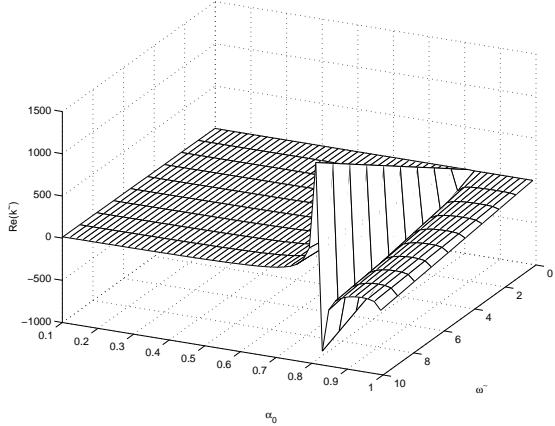}
\end{center}
\caption{\it Purely real longitudinal mode for the electron-positron plasma.}
\end{figure}
\subsubsection{Electron-ion Plasma}\label{subsubsec8.2.2}
In the case four modes are found to exist, two of which are purely real and others are complex conjugate modes. The mode shown in Fig. 1 for electron-positron plasma also found to exist for electron-ion plasma. The other real mode shown in Fig. 3 is a new mode for electron-ion plasma. Thus, for electron-ion plasma two purely real modes exist due to the difference in mass and density factors as between the positrons and ions. These two modes coalescence with a single mode for the special relativistic electron-positron plasma as investigated by SK \cite{twenty four}. The other two modes shown in Fig. 4, are a complex conjugate pair and are significantly damped and growing, respectively. The imaginary parts of this complex conjugate modes are equivalent to the two new purely imaginary modes we have found for electron-positron case discussed above with larger damping and growth rates. The differences in the magnitudes of the cyclotron frequencies $\omega _{c1}$ and $\omega _{c2}$ for the last two modes apparently lead to take the frequencies from their negative (and therefore unphysical) values for the electron-positron case to positive physical values for the electron-ion case. It is evident that these two modes are also stable and the damping and growth rates are independent of frequency, dependent only on the distance from the black hole horizon.

\subsection{High Frequency Transverse Modes}\label{subsec8.3}
\subsubsection{Electron-positron Plasma}\label{subsubsec8.3.1}
Four high frequency electromagnetic modes are found to exist for the electron-positron plasma. Two modes shown in Fig. 5 have same real parts as shown in Fig. 1. The effect of the general relativistic term is clearly shown here. For lower frequencies the plasma are unstable as moved toward the horizon and the damping and growth occurs as $\alpha\rightarrow 0$. It then appears, at a distance from the horizon corresponding to $\alpha_0<0.4$ and $\tilde \omega <6$ that energy is no longer fed into wave mode by the gravitational field but begins to be drained from the waves. The third mode, shown in Fig. 6, is also unstable near the event horizon. For lower frequencies the damping and growth rates are very high, but for higher frequencies the damped and growing rates are smaller. The fourth mode, shown in Fig. 7 is stable for $\alpha _0> 0.6$ and for higher frequencies, but unstable for $\alpha _0< 0.6$ and for lower frequencies. That is to say that, the damped and growing of these modes are depended on both the wave frequency and lapse function $\alpha _0$.
\subsubsection{Electron-ion Plasma}\label{subsubsec8.3.2}
Like the electron-positron plasma, the electron-ion plasma admits four high frequency modes. These are illustrated in Figs. 8 and 9. The Fig. 8 shows two complex modes having real parts equal and are similar with the modes shown in Fig. 1 for electron-positron plasma. Both the imaginary part of these two modes are stable for all frequencies and at all distances from the horizon corresponding to $\alpha_0>0.1$, but the solution becomes unstable for $\alpha_0<0.1$. This is also true for the remaining two modes shown in Fig. 9. The real part of this two modes are similar with the two real part of the modes shown in Figs. 6 and 7, respectively. Very closed to the horizon the growth and decay rates depend on both frequency and lapse function. The damping and growth are not clearly evident like the corresponding high frequency modes for electron-positron plasma.
\begin{figure}[h]\label{fig11}
\begin{center}
 \includegraphics[scale=.4]{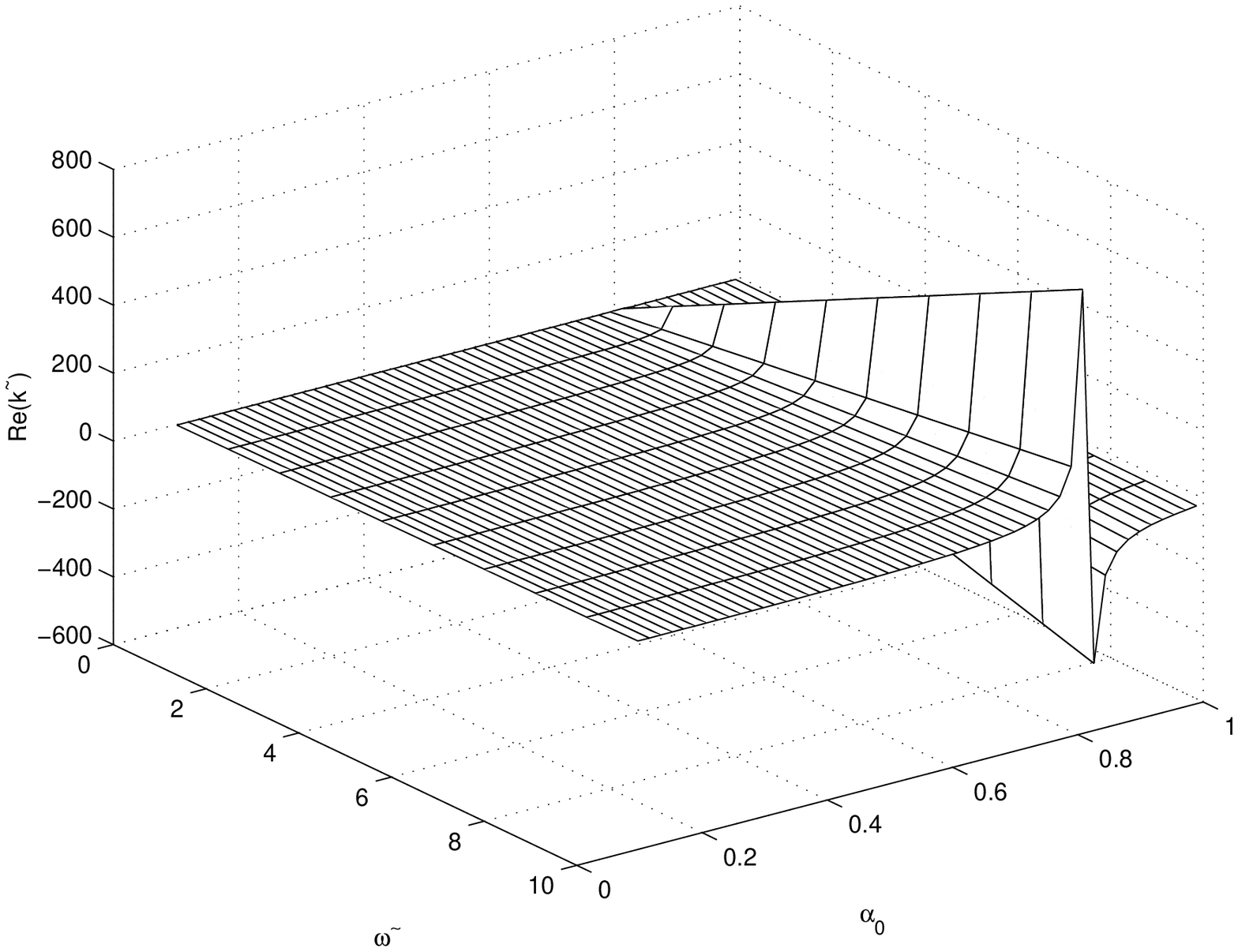}
 \includegraphics[scale=.4]{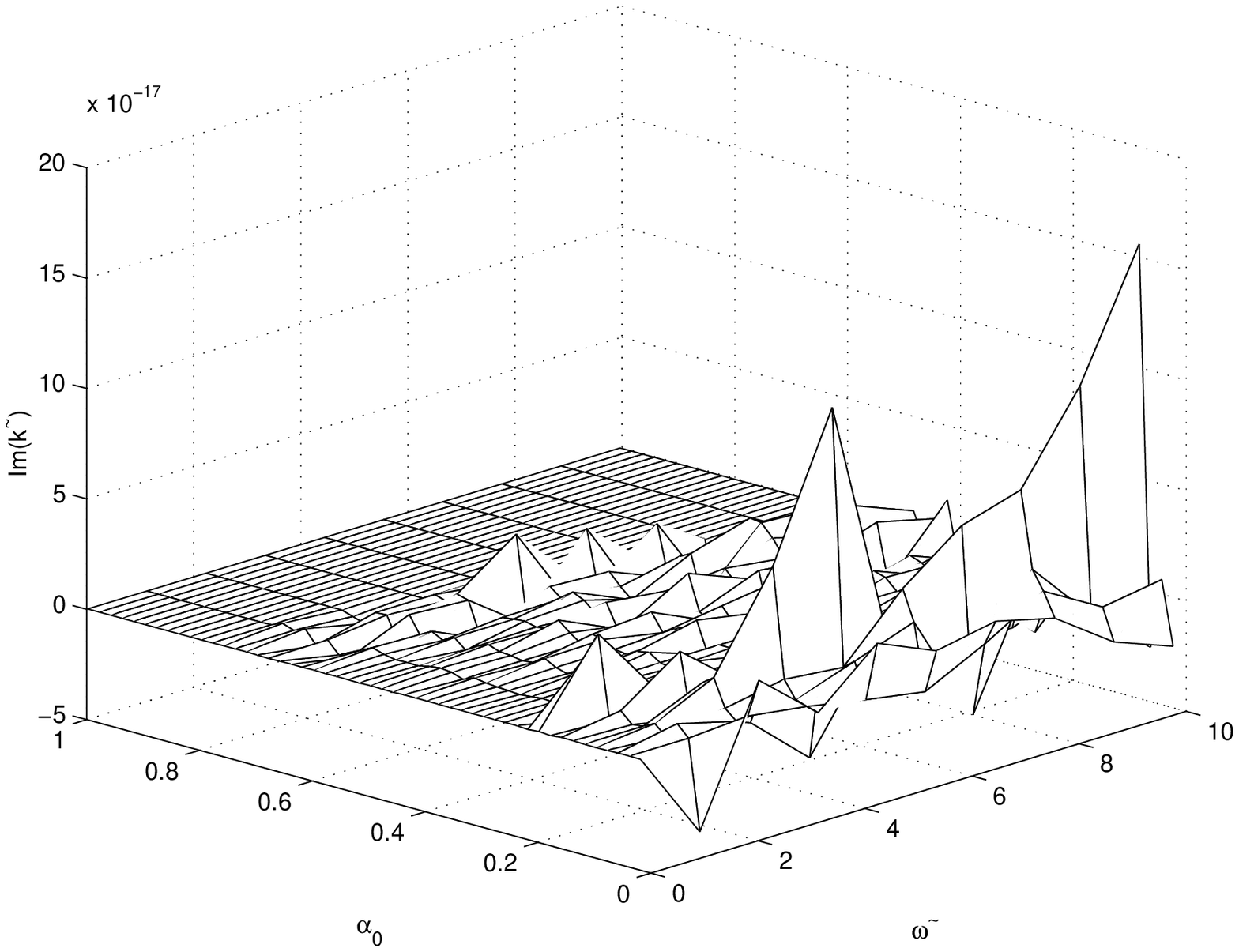}
\end{center}
\caption{\it {\bf Left}: Real part of a longitudinal complex mode for the electron-positron plasma. {\bf Right}: The imaginary part shows the interplay between frequency $\tilde\omega$ and $\alpha_0$ near the singularity and near the event horizon.}
\end{figure}
\begin{figure}[h]\label{fig12}
\begin{center}
 \includegraphics[scale=.4]{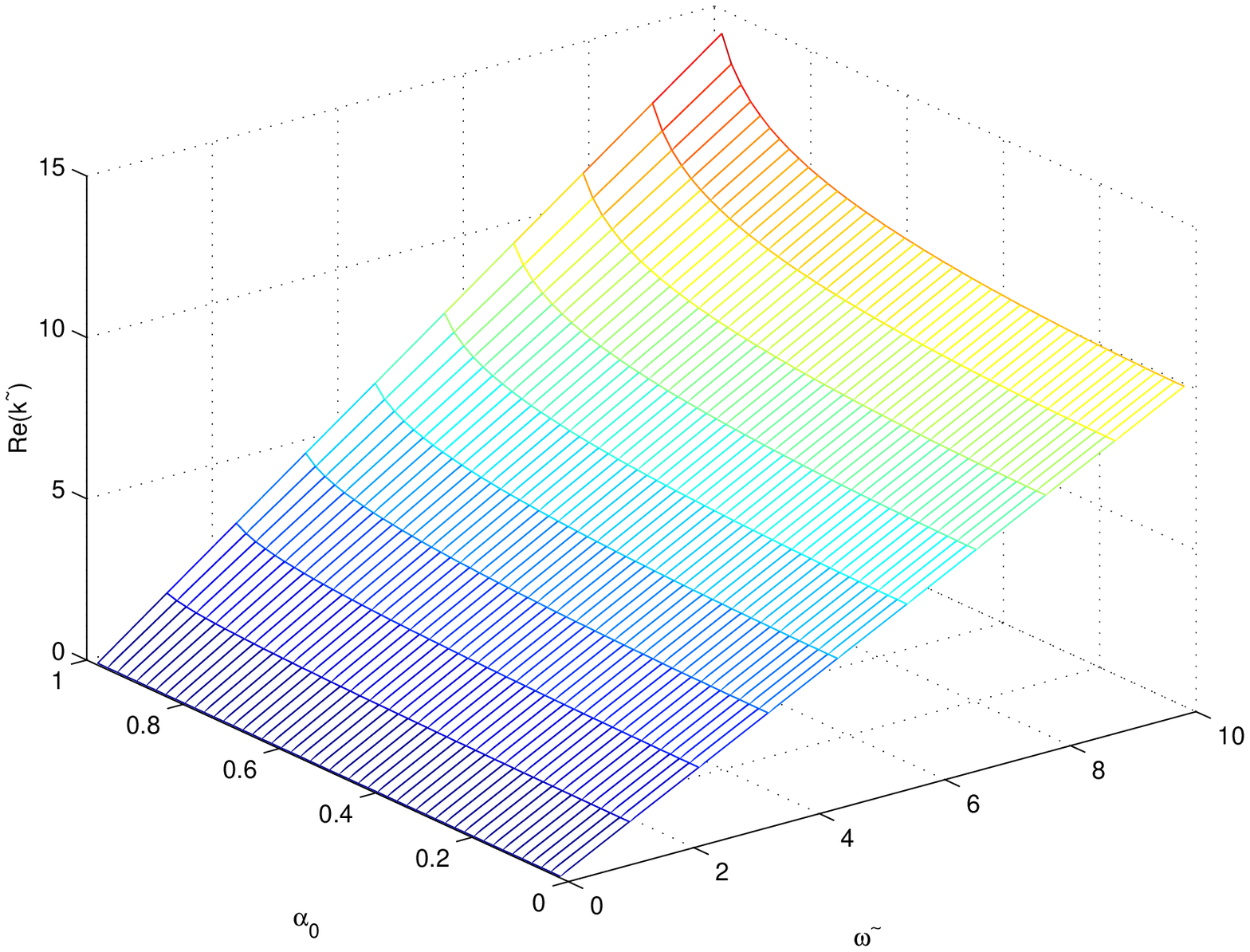}
 \includegraphics[scale=.4]{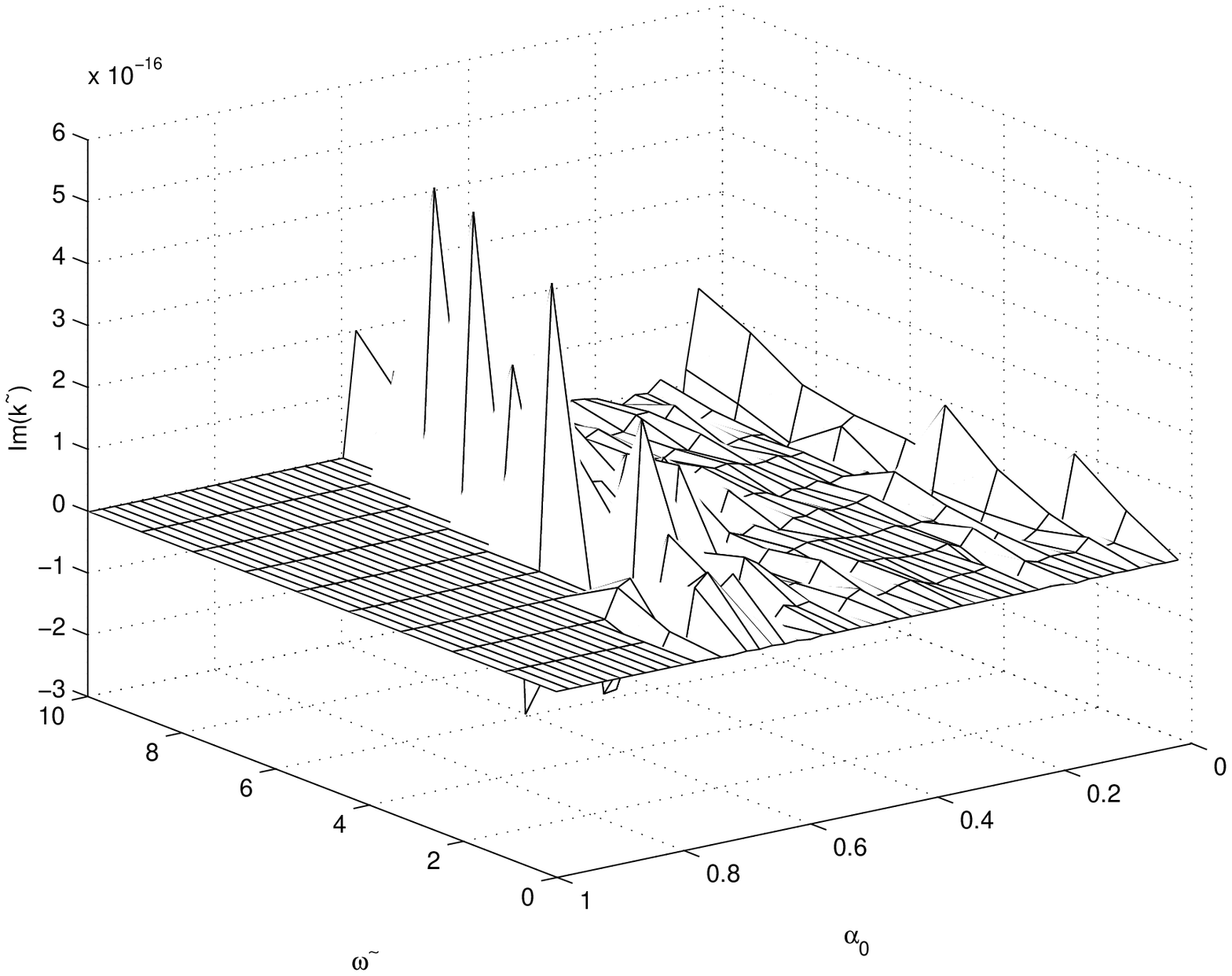}
\end{center}
\caption{\it {\bf Left}: Real part of a longitudinal complex mode for the electron-positron plasma. {\bf Right}: The imaginary part clearly shows the interplay between frequency $\tilde\omega$ and $\alpha_0$ near the transonic radius, $\alpha _t\sim 0.82$, and near the event horizon, $\alpha _0\rightarrow 0$.}
\end{figure}
\subsection{Longitudinal Modes}\label{subsec9.4}
\subsubsection{Electron-positron Plasma}\label{subsubsec9.4.1}
The longitudinal modes are not split into high and low frequency domain as was done for transverse electromagnetic waves because the same modes are exist in the low frequency, $0<\tilde\omega <1$, and high frequency, $1<\tilde\omega<10$, domains. So, $\tilde\omega$ has been chosen here as $0<\tilde\omega <10$. For electron-positron plasma, there exist five modes one of which is purely real mode shown in Fig. 10. This is also true for the special relativistic case investigated by SK where only one high frequency mode was found to exist for electron-positron plasma. For this mode transonic radius occurs at about $\alpha _t\sim 0.82$. The second mode shown in Fig. 11 is a complex mode with real part equal to the real modes shown above figure. In the imaginary part of this mode some interplay between frequency $\tilde\omega$ and $\alpha_0$ occur near the singularity and near the event horizon. The third modes shown in Fig. 12 also shows some interplay between $\tilde\omega$ and lapse function. The fourth mode shown in Fig. 13 is growth for  $\alpha _0<\alpha _t$ and damped for $\alpha _0>\alpha _t$. This mean that energy is drained from the wave rather than being fed into it by the gravitational field. The fifth mode is almost the opposite of the previous mode in that it is growth mode for $\alpha _0<\alpha _t$ and damped for $\alpha _0>\alpha _t$.
\subsubsection{Electron-ion Plasma}\label{subsubsec9.4.2}
As for electron-positron case, here five low frequency modes exist three of which are purely real and others are complex. Since the two-fluid (electron and ion) temperatures near the horizon are different, their transonic radii are different and are occurred at $\alpha _{t1}\sim 0.82$ and $\alpha _{t2}\sim 0.99$ respectively. Figure 14 shows one purely real mode for electron-positron plasma. The other two purely real modes are shown previously in Fig. 10 and the top of the Fig. 12. The fourth mode shown in Fig. 15, shows both damping and growth. It is stable for $\alpha _0<\alpha _{t1}$ and growth for $\alpha_{t1}<\alpha _0<\alpha _{t2}$ and then, for $\alpha _0>\alpha _{t2}$ it transition back to damped. The fifth mode, not shown here, is almost the opposite of the fourth mode and is stable for $\alpha _0<\alpha _{t1}$ and damped for $\alpha_{t1}<\alpha _0<\alpha _{t2}$ and then, for $\alpha _0>\alpha _{t2}$ it transition back to growth. These two modes also clearly show the influence of the transonic radii and again the fact that energy is being fed into the wave between the transonic radii but is drained from the wave very closed to the horizon.
\begin{figure}[h]\label{fig13}
\begin{center}
 \includegraphics[scale=.4]{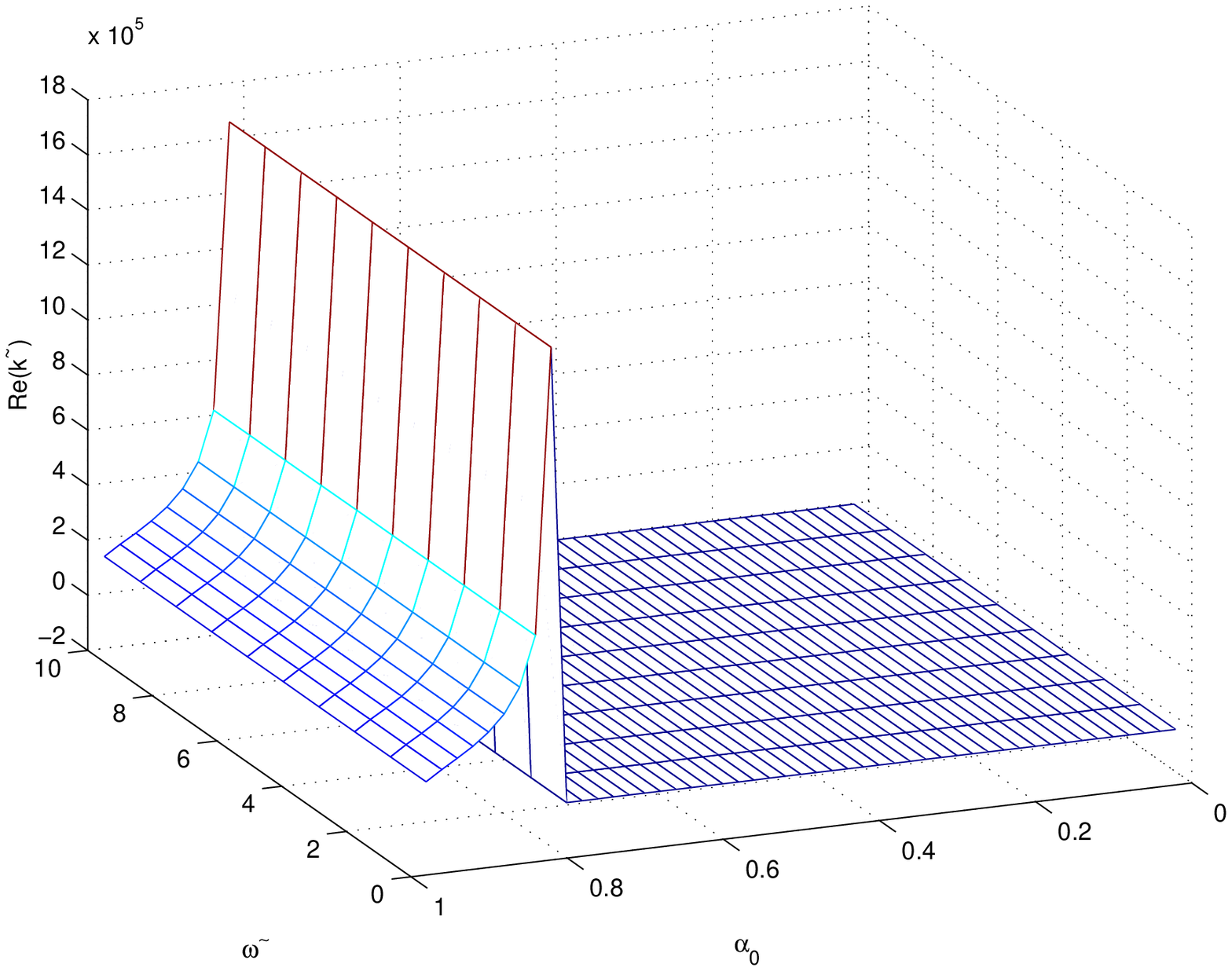}
 \includegraphics[scale=.4]{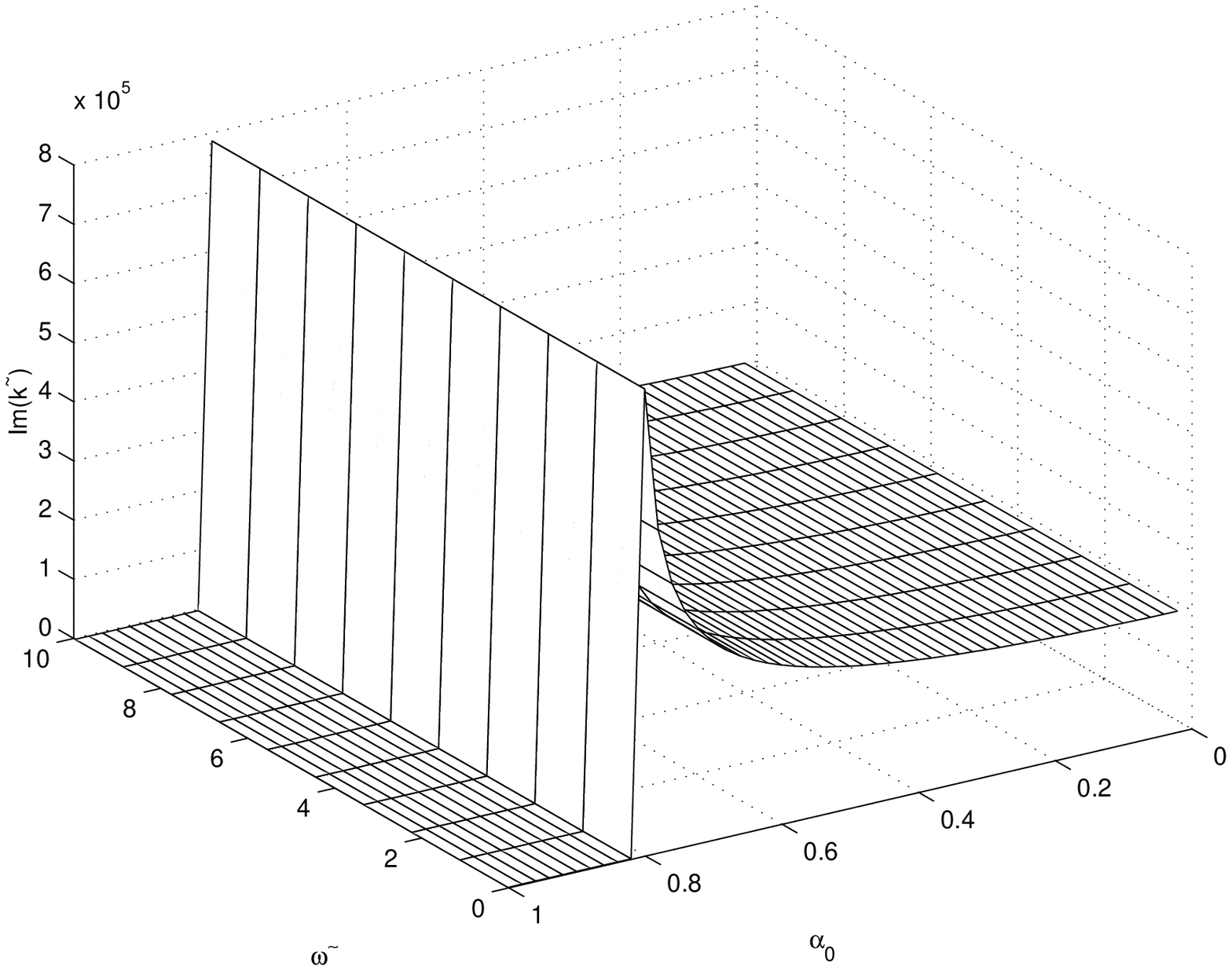}
\end{center}
\caption{\it {\bf Left}: Real part of a longitudinal complex mode for the electron-positron plasma. {\bf Right}: Imaginary part shows damped corresponding to $\alpha_0<\alpha _t$.}
\end{figure}
\begin{figure}[h]\label{fig14}
\begin{center}
   \includegraphics[scale=.4]{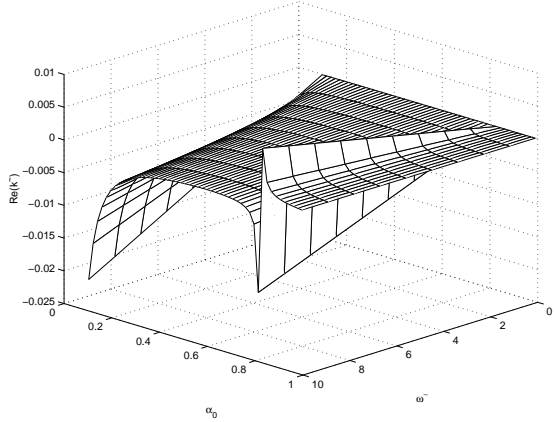}
\end{center}
\caption{\it Three purely real longitudinal modes exist for electron-ion plasma one of which is shown above and the another two are shown previously in Fig. 10 and the left of the Fig. 12.}
\end{figure}
\section{Concluding Remarks}\label{sec10}
The prime concern of this study has been exclusively the investigation, within the WKB approximation, of transverse electromagnetic and longitudinal waves in a two-fluid plasma closed to the event horizon of Schwarzschild black hole. We have derive the local dispersion relations for such types of waves and shown that the general relativistic effects enter these dispersion relations only in the ratio $\omega/\alpha_0$ and with $d\tau = \alpha_0 dt$ the transverse and longitudinal waves dispersion relations reduced to the special relativistic version as it should be according to the Einstein relativity principle. We solve the dispersion relations numerically for the wave number $k$. For the electron-positron plasma, the damping and growth rates are smaller in general, by several orders of magnitude, compared with the real components of the wave number. On the contrary, for the electron-ion plasma, modes with significant damping and growth rates are found to exist, in particular, for the Alfv\'en waves. Obviously, the damping and growth rates are frequency independent for the Alfv\'en waves, but they  are solely dependent on the radial distance from the horizon through the lapse function $\alpha _0$. However, this is not the case for the high frequency waves. The rate of damping or growth in that case is dependent on both frequency and radial distance from the horizon. These results are essentially agree with the results of SK \cite{twenty four} for ultrarelativistic limit and of Rahman \cite{thirty one} and Daniel and Tajima \cite{twenty three}for general relativity. For very low or negligible frequency, the plasmas are unstable, i.e., damped and growing modes exist as mansion above. Similar instabilities of a relativistic plasma were found by Mikhailovskii \cite{thirty three}, and Zaslavskii and Moiseev \cite{fourty}. The same conclusion also follows from the numerical calculations of Buzzi et al. \cite{twenty five, twenty six}.

For longitudinal waves the influence of transonic radius for each of the fluid species has been clearly evident. Like transverse electromagnetic waves damping and growth modes are found for each fluid species. The damped and growing rates for some of the modes dependent on frequency but for all the modes, the damping and growth rates are dependent on the radial distance from the horizon through the lapse function $\alpha _0$. For each fluid species, all the real transverse electromagnetic modes in the $\alpha_0$ domain are equivalent to all the real longitudinal modes in the $\alpha_0<\alpha_t$ domain and give the same results for the special relativistic case investigated by SK \cite{twenty four}. The presence of damped modes demonstrates that, at least in this approximation, energy is being drained from the waves by the gravitational field. Since the majority of the modes show growth rates, the  gravitational field is, in fact, feeding energy into the waves. This results are in accordance with the results of analytical studies carried out respectively by Mikhailovskii \cite{thirty three} and Rahman \cite{thirty one}.

\begin{figure}[h]\label{fig15}
\begin{center}
 \includegraphics[scale=.4]{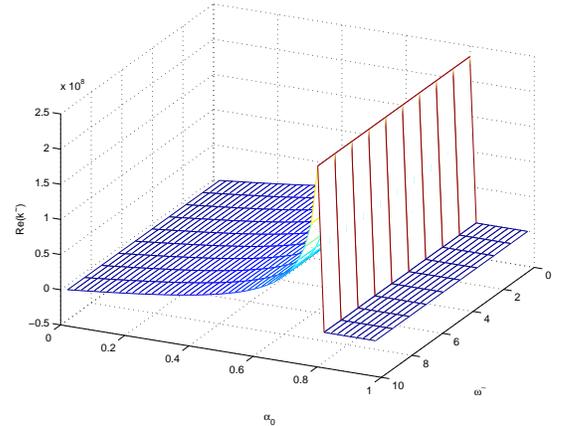}
 \includegraphics[scale=.4]{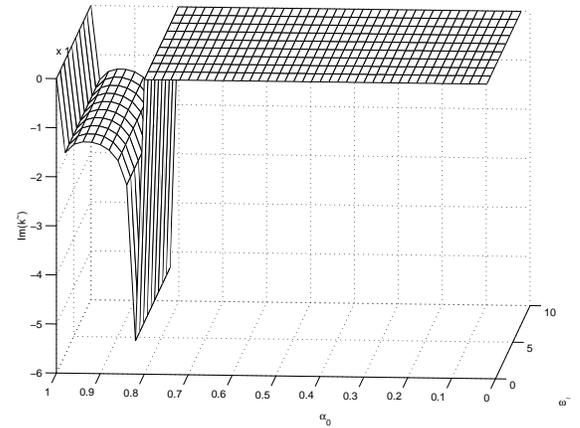}
\end{center}
\caption{\it {\bf Left}: Real part of a longitudinal complex mode for the electron-ion plasma. {\bf Right}: Imaginary part of longitudinal mode which is stable for $\alpha_0<\alpha _t$ and which shows growing and damped corresponding to $\alpha_{t1}<\alpha_0<\alpha _{t2}$ and $\alpha_0>\alpha_{t2}$, respectively. }
\end{figure}

\end{document}